\documentclass[12pt]{article}
\usepackage{amsfonts}
\usepackage{amsmath}
\usepackage{amssymb}
\usepackage{myart}

\normalbaselines
\input tcilatex
\begin{document}

\title{Physics geometrization in microcosm: discrete space-time and
relativity theory}
\author{Yuri A.Rylov}
\date{Institute for Problems in Mechanics, Russian Academy of Sciences,\\
101-1, Vernadskii Ave., Moscow, 119526, Russia.\\
e-mail: rylov@ipmnet.ru\\
Web site: {$http://rsfq1.physics.sunysb.edu/\symbol{126}rylov/yrylov.htm$}\\
or mirror Web site: {$http://gasdyn-ipm.ipmnet.ru/\symbol{126}%
rylov/yrylov.htm$}}
\maketitle

\begin{abstract}
The presented paper is a review of papers on the microcosm physics
geometrization in the last twenty years. These papers develop a new
direction of the microcosm physics. It is so-called geometric paradigm,
which is alternative to the quantum paradigm, which is conventionally used
now. The hypothesis on discreteness of the space-time geometry appears to be
more fundamental, than the hypothesis on quantum nature of microcosm.
Discrete space-time geometry admits one to describe quantum effects as pure
geometric effects. Mathematical technique of the microcosm physics
geometrization (geometric paradigm) is based on the physical geometry, which
is described completely by the world function. Equations, describing motion
of particles in the microcosm, are algebraic (not differential) equations.
They are written in a coordinateless form in terms of world function. The
geometric paradigm appeared as a result of overcoming of inconsistency of
the conventional elementary particle theory. In the suggested skeleton
conception the state of an elementary particle is described by its skeleton
(several space-time points). The skeleton contains all information on the
particle properties (mass, charge, spin, etc.). The skeleton conception is a
monistic construction, where elementary particle motion is described in
terms of skeleton and world function and only in these terms. The skeleton
conception can be constructed only on the basis of the physical geometry.
Unfortunately, most mathematicians do not accept the physical geometries,
because these geometries are nonaxiomatizable. It is a repetition of the
case, when mathematicians did not accept the non-Euclidean geometries of
Lobachevsky-Bolyai. As a result this review is a review of papers of one
author. This situation has some positive sides, because it appears to be
possible a consideration not only of papers, but also of motive for writing
some papers.
\end{abstract}

\section{Introduction}

The conventional paradigm of the microcosm physics development may be
classified as quantum paradigm. The quantum paradigm is based on hypothesis
of continuous space-time geometry equipped by quantum principles of particle
motion. There is an alternative geometric paradigm based on hypothesis on a
discrete space-time. There is no necessity to use quantum principles in the
geometric paradigm, because all quantum effects can be explained by
existence of elementary length of the discrete geometry. The elementary
length appears to be proportional to the quantum constant $\hbar $.

The hypothesis on discreteness of the space-time geometry looks more
reasonable and natural, than the hypothesis on mysterious quantum nature of
microcosm. One of reasons, why the geometric paradigm is not used in the
contemporary physics is the circumstance, that the discrete geometry has not
been developed properly. One believed that the discrete geometry is a
geometry on a lattice. Any lattice point set cannot be uniform and
isotropic, and such a point set is not adequate for the space-time.

In reality, the discrete space-time geometry can be defined on the same
point set, where the space-time geometry of Minkowski is given. In other
words, a discrete geometry may be uniform and isotropic. This unexpected
circumstance admits one to use a discrete geometry as a space-time geometry. 
\textit{The discrete geometry }$G_{\mathrm{d}}$\textit{\ is such a geometry,
where there are no close points.} Mathematically it means 
\begin{equation}
\left\vert \varrho \left( P,Q\right) \right\vert \notin \left( 0,\lambda
_{0}\right) ,\ \qquad \forall P,Q\in \Omega  \label{b1.1}
\end{equation}%
Here $\Omega $ is the point set, where the geometry is given, and $\varrho
\left( P,Q\right) $ is a distance between the points $P$, $Q$. The quantity $%
\lambda _{0}$ is the elementary length of the discrete geometry \ $\mathcal{G%
}_{\mathrm{d}}$. The geometry on a lattice can satisfy the property (\ref%
{b1.1}), but such a geometry cannot be uniform and isotropic. The discrete
space-time geometry has a set of new unexpected properties, which were
unknown in the twentieth century. This fact was one of reasons, why the
physics geometrization in microcosm has not been developed in the twentieth
century.

This paper is a short review of the physics geometrization development in
the last two decades. The physics geometrization began in the end of the
nineteenth century. Different stages of the physics geometrization are: (1)
connection of the conservation laws with the properties of the space-time
geometry (uniformity and isotropy), (2) the special relativity theory,(3)
the general relativity theory, (4) the Kaluza-Klein space-time geometry.
Most physicists do not believe in the physics geometrization in microcosm.
They believe in the quantum nature of physical phenomena in microcosm, and
they do not know properties of a discrete geometry, which admits one to
explain quantum phenomena as geometrical effects. It is a reason, why
practically nobody deal with the physics geometrization now. By necessity
this review of papers on the physics geometrization in microcosm is a review
of papers of one author.

It should note that we distinguish between a conception and a theory. A
conception does not coincide with a theory. For instance, the skeleton
conception of elementary particles distinguishes from a theory of elementary
particles. A conception investigates connections between concepts of a
theory. For instance, the skeleton conception of elementary particles
investigates the structure of a possible theory of elementary particles. It
investigates, why an elementary particle is described by its skeleton
(several space-time points), which contains all information on the
elementary particle. The skeleton conceptions explains, why dynamic
equations are coordinateless algebraic equations and why the dynamic
equations a written in terms of the world functions. However, the skeleton
conception does not answer the question, which skeleton corresponds to a
concrete elementary particle and what is the world function of the real
space-time. In other words, the skeleton conception deals with physical
principles, but not with concrete elementary particles. The conception
cannot be experimentally tested. However, if the world function of the real
space-time geometry has been determined and correspondence between a
concrete elementary particle and its skeleton has been established, the
skeleton conception turns to the elementary particle theory. The theory of
elementary particle (but not a conception) can be tested experimentally.

In other words, it is useless to speak on experimental test of the skeleton
conception, because it deals only with physical principles. Discussing
properties of a conception, one should discuss only properties of the
concept and logical connection between them, but not to what extent they
agree with experimental data.

We consider in the review the following problems

\begin{enumerate}
\item Conceptual defects of the quantum paradigm, which manifest themselves,
in particular, in incorrect use of the relativity principles at a
description of indeterministic particles.

\item Explanation of quantum effects as a statistical description of the
indeterministic particle motion.

\item Discrete geometry as a special case of a physical geometry and
properties of physical geometries.

\item Elementary particle dynamics in physical space-time geometry and
skeleton conception of particle dynamics.
\end{enumerate}

Idea of the physics geometrization is based on the following circumstance.
Description of the particle motion contains two essential elements: the
space-time geometry and the dynamic laws. The two categories are connected.
One can investigate the two categories only together, and the boundary
between the laws of geometry and the laws of dynamics is not fixed rigidly.
One can shift this boundary. For instance, one can choose a very simple
space-time geometry, then the laws of dynamics appear to be rather
complicated. One may try to use a complicated space-time geometry, which is
chosen in such a way, that the dynamic laws be very simple. For instance,
maybe, there exists such a space-time geometry, where the elementary
particles move freely. Interaction between particles is realized via the
space-time geometry. The Kaluza-Klein geometry is an example of such a
space-time geometry, where the electromagnetic field is a property of the
space-time geometry. If one uses the space-time geometry of Minkowski
(instead of the Kaluza-Klein geometry), the electromagnetic interaction of
particles is explained as a result of interaction with the electromagnetic
field,

\label{2b}The space-time of Minkowski is uniform and isotropic, and one can
easily write the conservation laws of energy-momentum and of angular
momentum in the Minkow\-ski space-time. One cannot write the conservation
laws in the Kaluza-Klein space-time with electromagnetic field, because this
space-time is not uniform and isotropic, in general. Such a difference is
conditioned by the circumstance, that in the space-time of Minkowski the
electromagnetic field is a substantive essence, whereas in the Kaluza-Klein
space-time the electromagnetic field is only a property of the space-time
geometry.

What point of view is true? We believe, that one should use both approaches.
In the geometrical approach the number of essences is less (in the limit of
a complete geometrization there is only one essence), and it is easier to
establish physical (and geometrical) principles responsible for description
of different sides of a physical phenomenon. On the other hand, when the
physical principles and connection between different sides of a physical
phenomenon have been established, one may consider the different sides of a
physical phenomenon as different essences. Such an approach admits one to
describe concrete physical phenomena easier and more convenient, considering
them as a result of interaction of different essences.

Developing the physics geometrization, one tries to work with physical
principles, assuming that the good old classical principles are true. We
stand aback from introducing new physical principles on the basis of
consideration of single physical phenomena. We believe that classical
physical principles are valid, although they are applied sometimes
incorrectly. We have succeeded to discover several mistakes in application
of classical principles of physics. Some of mistakes were connected with our
imperfect knowledge of geometry and, in particular, with imperfect knowledge
of a discrete geometry.\label{2e}

At the complete geometrization of physics the space-time geometry is chosen
in such a way, that all particles move freely. The force fields and their
interaction with particles appear only in the case, when the space-time
geometry is chosen incorrectly. In this case, when the chosen space-time
geometry differs from the true geometry, the deviation of geometries
generates appearance of force fields. The complete geometrization of physics
is known for classical (gravitational an electromagnetic) interactions .
However, it is not yet known in microcosm. The reason of this circumstance
lies mainly in the fact, that our knowledge of geometry is imperfect. The
complete geometrization of physics is possible only at a perfect description
of the space-time geometry.

A geometry as a science on disposition of geometrical objects in space or in
the event space (space-time) is described completely by the distance $\rho
\left( P,Q\right) $ between any two points $P$ and $Q$, or by the world
function $\sigma =\frac{1}{2}\rho ^{2}$. The geometry, which is described
completely by the world function will be referred to as a physical geometry.
After complete physics geometrization the particle dynamics turns to a
monistic conception, which is described completely in terms of one quantity
(world function). Any conception, which contains several basic concepts
(quantities), needs an agreement between all concepts, used in the
conception. Achievement of such an agreement is a very difficult problem.
One can see this in the example of a geometry. The physical geometry is a
monistic conception, because it is described by means of only world
function. One uses a few concepts (manifold, coordinate system, metric
tensor) in the conventional description of Riemannian geometries, and a
Riemannian geometry appears to be a less general conception, than a physical
geometry.

Albert Einstein dreamed on creation of a united field theory. Such a theory
was to be a monistic conception, and this circumstance was the most
attractive feature of such a theory. However, a monistic theory on the basis
of a geometry seems to be more attractive, than a monistic theory based on a
united field, because the main object of a physical geometry (world
function) is a simpler object, than a force field of the united field theory.

Problems of the physical geometrization appeared, when physicists began to
investigate physical phenomena in microcosm. We cannot know exactly the
microcosm space-time geometry. \ It is rather natural, that the space-time
geometry in microcosm may appear to be discrete. Contemporary researchers
consider a discrete geometry as a geometry on a lattice point set. In
particular, there is a special section in the ArXiv publications, entitled
High Energy Physics - Lattice. A lattice point set cannot be uniform and
isotropic. In accordance with this circumstance the discrete space-time
geometry (geometry on a lattice) is considered to be not uniform and
isotropic.

In reality a discrete space-time geometry is not a geometry on a lattice.
The discrete space-time geometry may be given on a continual point set. In
particular, it can be given on the same manifold, where the geometry of
Minkowski is given. It is connected with the fact, that the geometry
discreteness is a property of the geometry, but not a property of the point
set, where the geometry is given. A discrete geometry satisfies the
condition (\ref{b1.1}).

Geometry on a lattice satisfies the condition (\ref{b1.1}), but such a
geometry is a special kind of a discrete geometry, which cannot be uniform
and isotropic.

Let $\sigma _{\mathrm{M}}$ be the world function of the geometry Minkowski $%
\mathcal{G}_{\mathrm{M}}$%
\begin{equation}
\sigma _{\mathrm{M}}\left( x,x^{\prime }\right) =\frac{1}{2}g_{ik}\left(
x^{i}-x^{\prime i}\right) \left( x^{k}-x^{\prime k}\right) ,\qquad \sigma _{%
\mathrm{M}}\left( x,x^{\prime }\right) =\frac{1}{2}\rho _{\mathrm{M}%
}^{2}\left( x,x^{\prime }\right)  \label{b1.2}
\end{equation}%
where $\rho _{\mathrm{M}}\left( x,x^{\prime }\right) $ is the distance
(interval) between the points with inertial coordinates $x=\left\{
x^{0},x^{1},x^{2},x^{3}\right\} $ and $x^{\prime }=\left\{ x^{\prime
0},x^{\prime 1},x^{\prime 2},x^{\prime 3}\right\} $. The world function $%
\sigma _{\mathrm{d}}$%
\begin{equation}
\sigma _{\mathrm{d}}=\sigma _{\mathrm{M}}+\frac{\lambda _{0}^{2}}{2}\text{sgn%
}\left( \sigma _{\mathrm{M}}\right) ,\qquad \mathrm{sgn}\left( x\right)
=\left\{ 
\begin{array}{ccc}
1 & \text{if} & x>0 \\ 
0 & \text{if} & x=0 \\ 
-1 & \text{if} & x<0%
\end{array}%
\right.  \label{b1.3}
\end{equation}%
describes a discrete geometry $\mathcal{G}_{\mathrm{d}}$, which satisfies
the restriction (\ref{b1.1}), although the geometry $\mathcal{G}_{\mathrm{d}%
} $ is given on the same point set $\Omega _{\mathrm{M}}$, where the
geometry of Minkowski is given. The geometry $\mathcal{G}_{\mathrm{d}}$
appears to be uniform and isotropic.

\label{02b}However, one cannot use coordinates for description of the
geometry . It does not that one cannot introduce coordinates. Substituting (%
\ref{b1.2}) in (\ref{b1.3}), one obtains representation of the world
function $\sigma _{\mathrm{d}}$ in terms of coordinates. However, the
points, which have close coordinates, are not close in the sense that the
distance between them is greater, than $\lambda _{0}$%
\begin{equation}
\sqrt{2\sigma _{\mathrm{d}}\left( x,x^{\prime }\right) }\geq \lambda
_{0},\qquad 0<\left\vert x-x^{\prime }\right\vert ^{2}<\varepsilon
\label{b1.3a}
\end{equation}%
It means that coordinate lines and differentiation along them have no
relation to the discrete geometry $\mathcal{G}_{\mathrm{d}}$, given on the
manifold of Minkowski. It does not mean, that the discrete geometry $%
\mathcal{G}_{\mathrm{d}}$ does not exist. It means only, that capacities of
the coordinate description method are restricted, and one needs to use the
coordinateless method of description, which are used at description of
physical geometries \cite{R2000,R2002,R2004c}

\label{02e}Besides, the discrete geometry $\mathcal{G}_{\mathrm{d}}$ appears
to be multivariant and nonaxiomatizable \cite{R2006}. Such properties of a
geometry can be obtained only at a use of the coordinateless description
method. In the discrete space-time a particle cannot be described by a world
line, because any world line is a set of connected infinitesimal segments of
a straight line. However, in the discrete geometry $\mathcal{G}_{\mathrm{d}}$
there are no segments, whose length is shorter, than the elementary length $%
\lambda _{0}$. It means, that instead of world line one has a world chain%
\begin{equation}
\mathcal{C=}\dbigcup\limits_{s}\mathbf{P}_{s}\mathbf{P}_{s+1}  \label{b1.4}
\end{equation}%
consisting of geometrical vectors $\mathbf{P}_{s}\mathbf{P}_{s+1}=\left\{
P_{s},P_{s+1}\right\} $, $s=...-1,0,1,...$of finite length $\mu $. The
geometrical vector ($g$-vector) is an ordered set $\mathbf{PQ}=\left\{
P,Q\right\} $ of two points $P$ and $Q$. The first point $P$ is the origin
of the vector, whereas the second point $Q$ is the end of the $g$-vector.
Such a definition of the vector is used in physics. However, mathematicians
prefer another definition. They define a vector as an element of a linear
vector space.

\textit{Remark. }We used the special term "geometrical vector", because
conventionally the term " vector" means some many-component quantity
(components of the vector in some coordinate system). In general, a vector
is defined in the contemporary geometry as an element of the linear vector
space. In this case the vector can be decomposed over basic vectors of a
coordinate system and represented as a set of the vector coordinates. Such a
definition is convenient, when one speaks about vector field, having several
components. In the proper Euclidean geometry the concept of a geometrical
vector coincides with the conventional concept of a vector as an element of
the linear vector space. In the Euclidean geometry the $g$-vector can be
decomposed over basic vectors. It can be represented as a set of
coordinates. However, in the discrete geometry, described by the world
function (\ref{b1.3}), a geometric vector cannot be represented as a sum of
its projections onto basic vectors, because in the discrete geometry (\ref%
{b1.3}) one cannot introduce a linear vector space even locally. However,
the definition of a vector as a set of two points does not contain a
reference to a coordinate system and to special properties of the Euclidean
geometry (such as linear vector space). The definition of the geometrical
vector is more general, and according to the logic rules the term "vector"
should be used with respect to the geometric vector. Another term, for
instance, "linear vector" should be used for the vector, defined as an
element of the linear vector space.

The discrete geometry $\mathcal{G}_{\mathrm{d}}$ is obtained from the
geometry of Minkowski $\mathcal{G}_{\mathrm{M}}$ by means of a deformation
of the geometry of Minkowski, when the world function $\sigma _{\mathrm{M}}$
is replaced by the world function $\sigma _{\mathrm{d}}$ \cite{R2007}. World
chains in the discrete space-time geometry appear to be stochastic. Let the
elementary length $\lambda _{0}$ have the form%
\begin{equation}
\lambda _{0}^{2}=\frac{\hbar }{bc}  \label{b1.5}
\end{equation}%
where $\hbar $ is the quantum constant $c$ is the speed of the light, and $b$
is the universal constant, connecting the geometric mass $\mu $ (length of
the chain link) with the particle mass $m$ by means of%
\begin{equation}
m=b\mu  \label{b1.6}
\end{equation}%
Then statistical description of the stochastic world chains leads to the Schr%
\"{o}dinger equation \cite{R91}. As a result the quantum effects can be
described as geometrical effects of the discrete space-time geometry.
Quantum principles cease to be prime physical principles. They become
secondary principles, which should not be applied always and everywhere. In
particular, there is no necessity of the gravitational field quantization.

In the discrete space-time geometry the relativity theory appears to be
incomplete. The fact is that, the transition from the nonrelativistic
physics to the relativistic one is followed only by a modification of
dynamic equations, describing the particle motion. Description of the
particle state remains the same as in the nonrelativistic physics. The
particle state is described as a point in the phase space of coordinates and
momenta. The particle momentum $p_{k}$ is defined as a tangent vector to the
particle world line $x^{k}=x^{k}\left( \tau \right) $, $k=0,1,2,3$.

\begin{equation}
p_{k}\left( \tau \right) =\frac{mg_{kl}u^{l}\left( \tau \right) }{\sqrt{%
g_{js}u^{j}\left( \tau \right) u^{s}\left( \tau \right) }},\qquad
u^{l}\left( \tau \right) =\lim_{d\tau \rightarrow 0}\frac{x^{l}\left( \tau
+d\tau \right) -x^{l}\left( \tau \right) }{d\tau }  \label{b1.7}
\end{equation}%
where $\tau $ is a parameter along the world line. In the discrete
space-time geometry there are no world lines, and the limit (\ref{b1.7})
does not exist. This limit does not exist also in the case, when the
particle is indeterministic and its world line (if it exists) is random
(stochastic). In the physics of usual scale, when characteristic lengths
much more, than the elementary length $\lambda _{0}$, restricting the link
length of the world chain. In this case it is admissible to use the limit (%
\ref{b1.7}) as a good approximation. However, in the microcosm physics such
an approximation appears to be unsatisfactory, because characteristic
lengths of physical phenomena appear of the order of the elementary length $%
\lambda _{0}$. As a result the concepts of the elementary particle theory,
based on the particle state as point of the phase space appear to be
incomplete.

Consequent relativistic description of particles in microcosm does not use
the phase space and its points. Instead, one uses a skeleton conception of
the elementary particle description, where a particle is described by its
skeleton $\mathcal{P}_{n}=\left\{ P_{0},P_{1},...P_{n}\right\} $, which
consists of $n+1$ rigidly connected points $P_{0},P_{1},...P_{n}$. In the
case of pointlike particle its skeleton consists of two points $P_{0},P_{1}$%
, which define the particle momentum vector. In the given case all
characteristics of the particle (mass, charge, momentum) are defined
geometrically by the two points $P_{0},P_{1}$. In the case of a more
complicated particle, described by the skeleton $\mathcal{P}_{n}=\left\{
P_{0},P_{1},...P_{n}\right\} $, there are $n(n+1)/2$ invariants $\left\vert 
\mathbf{P}_{k}\mathbf{P}_{i}\right\vert $, $i,k=0,1,...n$, describing
geometrically all characteristics of the particle. The question about nature
of connection between the points of the skeleton does not arise, because the
discrete space-time geometry may have a restricted divisibility. Such a
question is conditioned by the hypothesis on continuous space-time geometry.

In the beginning of the twentieth century it was natural to think, that the
quantum particles are simply indeterministic (stochastic) particles,
something like Brownian particles. There were attempts to obtain quantum
mechanics as a statistical description of stochastically moving particles 
\cite{M49,F52}. However, these attempts failed, because a \textit{%
probabilistic conception of the statistical description} was used.

Statistical description is used in physics for description of
indeterministic particles (or systems), when there are no dynamic equations,
or initial conditions are indefinite. One considers statistical ensemble of
indeterministic particles, i.e. many independent similar particles. It
appears, that there are dynamic equations for the statistical ensemble $%
\mathcal{E}$ of indeterministic particles, although there are no dynamic
equations for a single indeterministic particle, which is a constituent of
this statistical ensemble $\mathcal{E}$. Consideration of the statistical
ensemble as a dynamic system is \textit{the}\emph{\ }\textit{dynamic
conception of the statistical description }(DCSD). It is a primordial
conception of statistical description. A use of DCSD is founded on
independence of constituents of the statistical ensemble. Random components
of motion are compensated due to their independence, whereas regular
components of motion are accumulated. As a result the statistical ensemble,
considered as a dynamic system, describes a mean motion of an
indeterministic particle.

In the nonrelativistic physics one uses the probabilistic conception of the
statistical description (PCSD). PCSD is used successfully, for instance, for
description of Brownian motion. In PCSD one traces the motion of a point in
the phase space. The point represents the state of indeterministic particle,
and motion of the point in the phase space is described by the transition
probability. Attempts of obtaining the quantum mechanics as a result of
statistical description in the framework PCSD failed \cite{M49,F52}, whereas
the statistical description in the framework of DCSD appeared to be
successful \cite{R71,R73a,R73}. This fact is explained by a use of the
dynamic conception of statistical description (DCSD), which does not use a
concept of the phase space.

In the relativistic case the ensemble state is described by a 4-vecotor $%
j^{k}\left( x\right) $, which described the density of world lines in the
vicinity of the point $x$. The ensemble state does not contain a reference
to the phase space. In the nonrelativistic case the ensemble state is
described by a 3-scalar $\rho \left( \mathbf{x},\mathbf{p}\right) $, which
describes the particle density in vicinity of the point $\left( \mathbf{x,p}%
\right) $ of the phase space. PCSD is based on a use of the nonnegative
quantity $\rho $, which is used as a probability density of the particle
position at the point $\left( \mathbf{x,p}\right) $ of the phase space.

Nonrelativistic quantum mechanics is a relativistic construction in reality,
because the stochastic component of the quantum particle motion may be
relativistic. At such a situation and one has to use the dynamic conception
of statistical description (DCSD), which does not use the nonrelativistic
concept of the phase space. Besides, one may not use the limit (\ref{b1.7})
in the definition for the particle momentum of stochastic world lines, which
can have no tangent vectors.

Indeed, in terms of DCSD one succeeded to obtain the quantum mechanics as a
statistical description of stochastically moving particles \cite%
{R71,R73a,R73}. This use of dynamic conception of statistical description
was not a stage of the physics geometrization. DCSD was simply an overcoming
of the incompleteness of the relativity theory, when relativistic dynamic
equations are combined with non-relativistic concept of the particle state.
However, the explanation of quantum mechanics effects as a result of
statistical description of stochastic particle motion arose the \textit{%
question on the nature of stochasticity of such a stochastic particle motion}%
.

Primarily the particle motion stochasticity was interpreted as a result of
interaction with an ether. However, further the idea has been appeared, that
the space-time geometry in itself may play the role of the ether. In other
words, the space-time geometry is to determine the free particle motion. If
the free particle motion is stochastic, the space-time geometry cannot be
geometry of Minkowski, because in the space-time geometry of Minkowski a
free particle motion is deterministic. The real space-time geometry is to be
uniform and isotropic, but it is to distinguish from the geometry of
Minkowski. It is to be multivariant. It means that at the point $Q_{0}$
there are many vectors $\mathbf{Q}_{0}\mathbf{Q}_{1}$, $\mathbf{Q}_{0}%
\mathbf{Q}_{1}^{\prime }$, $\mathbf{Q}_{0}\mathbf{Q}_{1}^{\prime \prime }$%
,..., which are equivalent to the vector $\mathbf{P}_{0}\mathbf{P}_{1}$ at
the point $P_{0}$. But vectors $\mathbf{Q}_{0}\mathbf{Q}_{1}$, $\mathbf{Q}%
_{0}\mathbf{Q}_{1}^{\prime }$,$\mathbf{Q}_{0}\mathbf{Q}_{1}^{\prime \prime }$%
,...are not equivalent between themselves. It means that the equivalence
relation is intransitive. Such a geometry cannot be axiomatizable, because 
\textit{in any axiomatizable geometry the equivalence relation is to be
transitive.} Nonaxiomatizable geometries were not known in seventieth of the
twentieth century. The discrete geometry (\ref{b1.3}) was not known also,
because in that time the discrete geometry was perceived as a geometry on a
lattice point set.

Idea of the physical geometry as a geometry described completely by the
world function appeared only in ninetieth of the twentieth century \cite%
{R1990}. The close idea of the distance (metrical) geometry appeared earlier 
\cite{M28,B53}. But such a geometry cannot be used as a space-time geometry.

One uses the discrete geometry (\ref{b1.3}) to explain the stochasticity of
free particle motion \cite{R91}. However, this geometry was used as a
simplest multivariant generalization of the geometry of Minkowski, but not
as a discrete space-time geometry. The fact, that the space-time geometry (%
\ref{b1.3}) is discrete, has been remarked several years later. \label{1b}
It is rather natural, that starting from idea of a discrete space-time
geometry, one comes to a geometry on a lattice, because one cannot obtain
the geometry (\ref{b1.3}), if concepts of physical geometry are unknown.%
\label{1e}

Application of a physical geometry for description of the space-time has
serious consequences for microcosm physics. It appears, that quantum
principles are not primary principles of nature. The relativity theory
appeared to be not completed. One needs to revise concept of the particle
state. The mathematical technique of description of the microcosm physical
phenomena changed essentially. Dynamic equations become finite difference
equations instead of differential equations. Description of particle motion
and that of gravitational field becomes coordinateless, and it was a
progress in the particle motion description.

Transition from the conventional description in terms of differential
equations to coordinateless description in terms of the world function
appears rather unexpected. It is connected with degenerative character of
the proper Euclidean geometry with respect to physical geometry. It means
that some geometrical concepts and some geometrical objects, which are
different in a physical geometry appear coinciding in the Euclidean
geometry. For instance, the geometrical vector $\mathbf{P}_{0}\mathbf{P}_{1}$
defined as the ordered set of two points $P_{0}$ and $P_{1}$ is a vector in
a physical geometry and in the Euclidean geometry. Projections $p_{l}$ of
vector $\mathbf{P}_{0}\mathbf{P}_{1}$ on basic coordinate vectors $\mathbf{Q}%
_{0}\mathbf{O}_{l}$, $l=1,2,..n$ are defined by the relation%
\begin{equation}
p_{l}=\left( \mathbf{P}_{0}\mathbf{P}_{1}.\mathbf{Q}_{0}\mathbf{Q}%
_{l}\right) ,\qquad l=1,2,...3  \label{b1.8}
\end{equation}%
Here $\left( \mathbf{P}_{0}\mathbf{P}_{1}.\mathbf{Q}_{0}\mathbf{Q}%
_{l}\right) $ is the scalar product of two vectors $\mathbf{P}_{0}\mathbf{P}%
_{1}$ and $\mathbf{Q}_{0}\mathbf{Q}_{l}$, which is defined in terms of the
world function by the relation 
\begin{equation}
\left( \mathbf{P}_{0}\mathbf{P}_{1}.\mathbf{Q}_{0}\mathbf{Q}_{l}\right)
=\sigma \left( P_{0},Q_{l}\right) +\sigma \left( P_{0},Q_{l}\right) -\sigma
\left( P_{0},Q_{l}\right) -\sigma \left( P_{0},Q_{l}\right)  \label{b1.9}
\end{equation}%
The expression of the scalar product (\ref{b1.9}) via the world function is
the same in a physical geometry and in the proper Euclidean one. In the
physical geometry the relation (\ref{b1.9}) is a definition of the scalar
product, whereas in the Euclidean geometry the relation (\ref{b1.9}) is
obtained as a corollary of the cosine theorem, but in both cases the
expression (\ref{b1.9}) is true. The scalar product has conventional linear
properties in the Euclidean geometry, but these properties are absent, in
general, in the physical geometry. As a result components $p_{l}$ of the
geometrical vector $\mathbf{P}_{0}\mathbf{P}_{1}$ do not determine the
vector $\mathbf{P}_{0}\mathbf{P}_{1}$ in the physical geometry, although
they determine the vector $\mathbf{P}_{0}\mathbf{P}_{1}$ in the proper
Euclidean geometry. It means, that the vector $\mathbf{P}_{0}\mathbf{P}_{1}$
and its components $p_{l}$, $l=1,2,..n$ mean the same quantity in the
Euclidean geometry, whereas they are, in general, different quantities in a
physical geometry.

In a like way the expression for a circular cylinder $Cyl_{P_{0}P_{1}Q}$,
determined by points $P_{0},P_{1}$ $\left( P_{0}\neq P_{1}\right) $ on the
cylinder axis and by the point $Q$ on cylinder surface, is a set of points $%
R $, satisfying the relation 
\begin{equation}
Cyl_{P_{0}P_{1}Q}=\left\{ R|S_{P_{0}P_{1}R}=S_{P_{0}P_{1}Q}\right\}
\label{b1.10}
\end{equation}%
where $S_{P_{0}P_{1}Q}$ $\ $is the area of the triangle, determined by
vertices $P_{0},P_{1},Q$. The area $S_{P_{0}P_{1}Q}$ is calculated by means
of the Heron formula via distances between the points $P_{0},P_{1},Q$. Let
the point $P_{3}\in \mathcal{T}_{\left[ P_{0}P_{1}\right] }$, where $%
\mathcal{T}_{\left[ P_{0}P_{1}\right] }$ is a segment of a straight line
between the points $P_{0},P_{1}$. This segment is defined as a set of points 
$R$ by the relation 
\begin{equation}
\mathcal{T}_{\left[ P_{0}P_{1}\right] }=\left\{ R|\sqrt{2\sigma \left(
P_{0},R\right) }+\sqrt{2\sigma \left( P_{1},R\right) }=\sqrt{2\sigma \left(
P_{0},P_{1}\right) }\right\}  \label{b1.11}
\end{equation}%
Then in the proper Euclidean geometry $%
Cyl_{P_{0}P_{1}Q}=Cyl_{P_{0}P_{3}Q}=Cyl_{P_{1}P_{3}Q}$. However, in a
physical geometry, in general, $Cyl_{P_{0}P_{1}Q}\neq Cyl_{P_{0}P_{3}Q}\neq
Cyl_{P_{1}P_{3}Q}$. In other words, many different cylinders $%
Cyl_{P_{0}P_{1}Q},\ \ \ P_{0},P_{1}\in \mathcal{T}_{\left[ S_{1}S_{2}\right]
}$ of a physical geometry degenerate in the proper Euclidean geometry into
one cylinder, defined by its axis $\mathcal{T}_{\left[ S_{1}S_{2}\right] }$
and by the point $Q$ on the surface of the cylinder. This fact takes place,
because the segment of the straight line (\ref{b1.11}) is one-dimensional in
the case of the proper Euclidean geometry, but it is, in general, a
many-dimensional surface in the case of a physical geometry.

One-dimensionality of $\mathcal{T}_{\left[ S_{1}S_{2}\right] }$ in the
Euclidean geometry is formulated in terms of the world function as follows.
Any section $S\left( \mathcal{T}_{\left[ S_{1}S_{2}\right] },Q\right) $ of
the segment $\mathcal{T}_{\left[ S_{1}S_{2}\right] }$ at the point $Q\in 
\mathcal{T}_{\left[ S_{1}S_{2}\right] }$ consists of one point $Q$. Section $%
S\left( \mathcal{T}_{\left[ S_{1}S_{2}\right] },Q\right) $ is defined as a
set of points $R$ 
\begin{equation}
S\left( \mathcal{T}_{\left[ S_{1}S_{2}\right] },Q\right) =\left\{ R|\sigma
\left( S_{1},R\right) =\sigma \left( S_{1},Q\right) \wedge \sigma \left(
S_{2},R\right) =\sigma \left( S_{2},Q\right) \right\}  \label{b1.12}
\end{equation}%
In the proper Euclidean geometry $S\left( \mathcal{T}_{\left[ S_{1}S_{2}%
\right] },Q\right) =\left\{ Q\right\} ,$ $\forall Q\in \mathcal{T}_{\left[
S_{1}S_{2}\right] }$, whereas in the case of a physical geometry this
equality does not take place, in general.

Thus, the physical geometry degenerates, in general, at a transition from a
physical geometry to the proper Euclidean geometry. Different geometrical
objects and concepts may coincide. On the contrary at transition from the
proper Euclidean geometry to a physical geometry some geometrical objects
split into different geometrical objects. Transition from a general case to
a special one, followed by a degeneration, is perceived easily, whereas a
transition from a special case to a general one, followed by a splitting of
geometrical objects and of geometrical concepts, is perceived hard.

\section{Relativistic invariance}

The relativistic invariance is presented usually as an invariance of dynamic
equation with respect to the Poincare group of inertial coordinate
transformations. Nonrelativistic dynamic equations are considered to be
invariant with respect to Galilean group of inertial coordinates
transformation. Is it possible to formulate difference between relativistic
physics and nonrelativistic one in invariant terms, i.e. without a reference
to coordinate system and the laws of their transformation? Yes, it is
possible.

In the relativistic physics the space-time geometry is described by means of
one structure $\sigma $, which is known as the squared space-time interval,
or the world function. In the nonrelativistic physics the event space
(space-time) is described by two invariant geometrical structures. Such a
two-structure description is not a space-time geometry, because the
space-time geometry is described by one structure $\sigma $. If there exist
another space-time structure, such a construction should be referred to as a
fortified geometry, i.e. a geometry with additional geometric structure.
This additional structure is the time structure $T\left( P,Q\right) $ which
is a difference of absolute times between the points $P$ and $Q$. One can
construct another geometrical structure $S\left( P,Q\right) $, which is a
difference between of absolute spatial positions of points $P$ and $Q$. The
structure $S\left( P,Q\right) $ is not an independent structure. The spatial
structure $S\left( P,Q\right) $ can be constructed of two structures $\sigma 
$ and $T$. In any case \textit{in the nonrelativistic physics there are two
independent geometrical structures}. \textit{In the relativistic physics
there is only one structure }$\sigma $.

Usually one uses the time structure $T$ and the spatial structure $S$ in the
nonrelativistic physics. However, one may use geometrical structures $\sigma 
$ and $T$. In this case one can investigate additional restrictions, imposed
by time structure $T$ on the space-time geometry of Minkowski. Geometrical
structures of the space-time determine a motion group of the space-time, and
this motion group determines group of invariance of dynamic equations. Thus,
the difference between the relativistic physics and nonrelativistic one is
determined by the number of geometrical structures. This difference may be
formulated in coordinateless form. The transformation laws of dynamic
equations are only corollaries of these geometrical structures existence.

\section{Statistical description of the stochastic particle motion}

Statistical description of stochastic (indeterministic) particles was an
origin of the physical geometry, because, it put the question on a nature of
this indeterminism, which can be explained only by a more general uniform
space-time geometry, than the geometry of Minkowski.

As we have mentioned in the introduction, a statistical description of
indeterministic particles was made at first by means of the dynamical
conception of statistical description (DCSD). This approach is founded on a
use of relativistic concept of particle state \cite{R71,R73a,R73}. Another
method of the stochastic particles description has been used later, when a
statistical ensemble (instead of a single particle) has been considered as a
basic element of the particle dynamics \cite{R2006a}. The concept of a
single particle and the concept of the phase space are not used in this
method. This method goes around the nonrelativistic concept of the particle
state. It does not use the concept of the particle state. It uses only
concept of the ensemble state, which is insensitive to the problem of the
limit (\ref{b1.7}) existence. From formal viewpoint this method uses DCSD,
but not PCSD.

The action for the statistical ensemble $\mathcal{E}\left[ \mathcal{S}_{%
\mathrm{st}}\right] $ of free indeterministic particles $\mathcal{S}_{%
\mathrm{st}}$ is written in the form%
\begin{equation}
\mathcal{A}_{\mathcal{E}\left[ \mathcal{S}_{\mathrm{st}}\right] }\left[ 
\mathbf{x},\mathbf{u}\right] =\int \dint\limits_{V_{\xi }}\left\{ \frac{m}{2}%
\mathbf{\dot{x}}^{2}+\frac{m}{2}\mathbf{u}^{2}-\frac{\hbar }{2}\mathbf{%
\nabla u}\right\} dtd\mathbf{\xi },\qquad \mathbf{\dot{x}\equiv }\frac{d%
\mathbf{x}}{dt}  \label{d1.5}
\end{equation}%
Independent variables $\mathbf{\xi }=\left\{ \xi _{1},\xi _{2},\xi
_{3}\right\} $ label constituents $\mathcal{S}_{\mathrm{st}}$ of the
statistical ensemble. The dependent variable $\mathbf{x}=\mathbf{x}\left( t,%
\mathbf{\xi }\right) $ describes the regular component of the particle
motion. The variable $\mathbf{u}=\mathbf{u}\left( t,\mathbf{x}\right) $
describes the mean value of the stochastic velocity component, $\hbar $ is
the quantum constant. The second term in (\ref{d1.5}) describes the kinetic
energy of the stochastic velocity component. The third term describes
interaction between the stochastic component $\mathbf{u}\left( t,\mathbf{x}%
\right) $ and the regular component $\mathbf{\dot{x}}\left( t,\mathbf{\xi }%
\right) $. The operator 
\begin{equation}
\mathbf{\nabla =}\left\{ \frac{\partial }{\partial x^{1}},\frac{\partial }{%
\partial x^{2}},\frac{\partial }{\partial x^{3}}\right\}  \label{d1.5a}
\end{equation}%
is defined in the space of coordinates $\mathbf{x}$. Dynamic equations for
the dynamic system $\mathcal{E}\left[ \mathcal{S}_{\mathrm{st}}\right] $ are
obtained as a result of variation of the action (\ref{d1.5}) with respect to
dynamic variables $\mathbf{x}$ and $\mathbf{u}$.

The action for a single indeterministic particle $\mathcal{S}_{\mathrm{st}}$
has the form%
\begin{equation}
\mathcal{A}_{\mathcal{S}_{\mathrm{st}}}\left[ \mathbf{x},\mathbf{u}\right]
=\int \dint\limits_{V_{\xi }}\left\{ \frac{m}{2}\mathbf{\dot{x}}^{2}+\frac{m%
}{2}\mathbf{u}^{2}-\frac{\hbar }{2}\mathbf{\nabla u}\right\} dt,\qquad 
\mathbf{\dot{x}\equiv }\frac{d\mathbf{x}}{dt}  \label{a1.11}
\end{equation}%
This action is not correctly defined, because operator $\mathbf{\nabla }$ is
defined on 3D-space of coordinates $\mathbf{x}=\left\{
x^{1},x^{2},x^{3}\right\} $, whereas in the action functional (\ref{a1.11})
the variable $\mathbf{x}$ is used only on one-dimensional set. It means that
there are no dynamic equations for the particle $\mathcal{S}_{\mathrm{st}}$,
and the particle $\mathcal{S}_{\mathrm{st}}$ is a stochastic
(indeterministic) system. However, the action functional (\ref{d1.5}) is
well defined, and dynamic equations exist for the statistical ensemble $%
\mathcal{E}\left[ \mathcal{S}_{\mathrm{st}}\right] $, although dynamic
equations do not exist for constituents of this statistical ensemble.

Variation of the action (\ref{d1.5}) leads to dynamic equations%
\begin{equation}
\delta \mathbf{u:\qquad }m\rho \mathbf{u}+\frac{\hbar }{2}\mathbf{\nabla }%
\rho =0,\qquad \mathbf{u}=-\frac{\hbar }{2m}\mathbf{\nabla }\ln \rho
\label{d1.7}
\end{equation}%
\begin{equation}
\delta \mathbf{x:\qquad }m\frac{d^{2}\mathbf{x}}{dt^{2}}=\mathbf{\nabla }%
\left( \frac{m}{2}\mathbf{u}^{2}-\frac{\hbar }{2}\mathbf{\nabla u}\right)
\label{d1.8}
\end{equation}%
where%
\begin{equation}
\rho =\frac{\partial \left( \xi _{1},\xi _{2},\xi _{3}\right) }{\partial
\left( x^{1},x^{2},x^{3}\right) }=\left( \frac{\partial \left(
x^{1},x^{2},x^{3}\right) }{\partial \left( \xi _{1},\xi _{2},\xi _{3}\right) 
}\right) ^{-1}  \label{d1.7a}
\end{equation}

After proper change of variables the dynamic equations are reduced to the
equation \cite{R2006a}%
\begin{equation}
i\hbar \partial _{0}\psi +\frac{\hbar ^{2}}{2m}\mathbf{\nabla }^{2}\psi +%
\frac{\hbar ^{2}}{8m}\mathbf{\nabla }^{2}s_{\alpha }\cdot \left( s_{\alpha
}-2\sigma _{\alpha }\right) \psi -\frac{\hbar ^{2}}{4m}\frac{\mathbf{\nabla }%
\rho }{\rho }\mathbf{\nabla }s_{\alpha }\sigma _{\alpha }\psi =0
\label{d6.14}
\end{equation}%
where $\psi $ is the two-component complex wave function%
\begin{equation}
\rho =\psi ^{\ast }\psi ,\qquad s_{\alpha }=\frac{\psi ^{\ast }\sigma
_{\alpha }\psi }{\rho },\qquad \alpha =1,2,3  \label{d6.7}
\end{equation}%
$\sigma _{\alpha }$ are $2\times 2$ Pauli matrices%
\begin{equation}
\sigma _{1}=\left( 
\begin{array}{cc}
0 & 1 \\ 
1 & 0%
\end{array}%
\right) ,\qquad \sigma _{2}=\left( 
\begin{array}{cc}
0 & -i \\ 
i & 0%
\end{array}%
\right) ,\qquad \sigma _{3}=\left( 
\begin{array}{cc}
1 & 0 \\ 
0 & -1%
\end{array}%
\right) ,  \label{d6.8}
\end{equation}

If components $\psi _{1}$ and $\psi _{2}$ are linear dependent $\psi =\left( 
\begin{array}{c}
a\psi _{1} \\ 
b\psi _{1}%
\end{array}%
\right) $, $a,b=\mathrm{const}$, then $\mathbf{\ s}=\mathrm{const}$. Two
last terms of the equation (\ref{d6.14}) vanish, and the equation turns to
the Schr\"{o}dinger equation%
\begin{equation}
i\hbar \partial _{0}\psi +\frac{\hbar ^{2}}{2m}\mathbf{\nabla }^{2}\psi =0
\label{a1.12}
\end{equation}

Thus, the Schr\"{o}dinger equation and interpretation of the quantum
mechanics appear from the dynamical system $\mathcal{E}\left[ \mathcal{S}_{%
\mathrm{st}}\right] $, described by the action functional (\ref{d1.5}). This
fact seems rather unexpected, because in quantum mechanics the wave function
is considered as a specific quantum object, which has no analog in classical
physics. In reality, the wave function is simply a way of description of
ideal continuous medium \cite{R1999}. One may describe an ideal fluid in
terms of hydrodynamic variables: density $\rho $ and velocity $\mathbf{v}$%
\textbf{. }One may describe an ideal fluid in terms of the wave function.
The statistical ensemble $\mathcal{E}\left[ \mathcal{S}_{\mathrm{st}}\right] 
$ is a dynamic system of the type of continuous medium. The two
representations of dynamic equations for the dynamic system $\mathcal{E}%
\left[ \mathcal{S}_{\mathrm{st}}\right] $ can be transformed one into
another.

Generalization of the action (\ref{a1.11}) on the stochastic relativistic
charged particle, moving in an electromagnetic field, has the form \cite%
{R2003} 
\begin{eqnarray}
\mathcal{A}\left[ x,\kappa \right] &=&\int \left\{ -mcK\sqrt{g_{ik}\dot{x}%
^{i}\dot{x}^{k}}-\frac{e}{c}A_{k}\dot{x}^{k}\right\} d^{4}\xi ,\qquad
d^{4}\xi =d\xi _{0}d\mathbf{\xi ,}  \label{A.1} \\
K &=&\sqrt{1+\lambda ^{2}\left( \kappa _{l}\kappa ^{l}+\partial _{l}\kappa
^{l}\right) },\qquad \lambda =\frac{\hbar }{mc}  \label{A.2}
\end{eqnarray}%
where $x=\left\{ x^{i}\left( \xi _{0},\mathbf{\xi }\right) \right\} ,\;$ $%
i=0,1,2,3$ are dependent variables. $\xi =\left\{ \xi _{0},\mathbf{\xi }%
\right\} =\left\{ \xi _{k}\right\} ,\;\;k=0,1,2,3$ are independent
variables, and $\dot{x}^{i}\equiv dx^{i}/d\xi _{0}.$ The quantities $\kappa
^{l}=\left\{ \kappa ^{l}\left( x\right) \right\} ,\;$ $l=0,1,2,3$ are
dependent variables, describing stochastic component of the particle motion, 
$A_{k}=\left\{ A_{k}\left( x\right) \right\} ,\;\;k=0,1,2,3$ is the
potential of electromagnetic field. The dynamic system, described by the
action (\ref{A.1}), (\ref{A.2}) is a statistical ensemble of indeterministic
particles, which looks as some continuous medium. The variables $\kappa ^{l}$
are connected with the stochastic component $u^{l}$ of the particle
4-velocity by the relation%
\begin{equation}
\kappa ^{l}=\frac{m}{\hbar }u^{l},\qquad l=0,1,2,3  \label{b3.1}
\end{equation}%
In the nonrelativistic approximation one may neglect the temporal component $%
\kappa ^{0}=\frac{m}{\hbar }u^{0}$ with respect to the spatial one $\mathbf{%
\kappa }=\frac{m}{\hbar }\mathbf{u}.$ Setting $\xi _{0}=t$ $=x^{0}$ and $%
A_{k}=0$ in (\ref{A.1}), (\ref{A.2}), we obtain the action (\ref{d1.5})
instead of (\ref{A.1}), (\ref{A.2}).

After a proper change of variables one obtains dynamic equation for the
action (\ref{A.1}), (\ref{A.2}). This dynamic equation has the form \cite%
{R2003} 
\begin{eqnarray}
&&\left( -i\hbar \partial _{k}+\frac{e}{c}A_{k}\right) \left( -i\hbar
\partial ^{k}+\frac{e}{c}A^{k}\right) \psi -\left( m^{2}c^{2}+\frac{\hbar
^{2}}{4}\left( \partial _{l}s_{\alpha }\right) \left( \partial ^{l}s_{\alpha
}\right) \right) \psi  \notag \\
&=&-\hbar ^{2}\frac{\partial _{l}\left( \rho \partial ^{l}s_{\alpha }\right) 
}{2\rho }\left( \sigma _{\alpha }-s_{\alpha }\right) \psi  \label{b3.3}
\end{eqnarray}%
where designations (\ref{d6.7}), (\ref{d6.8}) are used. In the case, when
the wave function $\psi $ is one-component, vector $\mathbf{s=}$const, and
the dynamic equation (\ref{b3.3}) turns to the Klein-Gordon equation%
\begin{equation}
\left( -i\hbar \partial _{k}+\frac{e}{c}A_{k}\right) \left( -i\hbar \partial
^{k}+\frac{e}{c}A^{k}\right) \psi -m^{2}c^{2}\psi =0  \label{b3.4}
\end{equation}

Transformation of hydrodynamic equations (\ref{d1.7}) into dynamic equations
in terms of the wave function $\psi $ is based on the fact, that a wave
function is a method of description of hydrodynamic equations \cite{R1999}.
Transformation of hydrodynamic equations, described in terms of hydrodynamic
variables (density $\rho $ and velocity $\mathbf{v}$\textbf{), }to a
description in terms the wave function rather is bulky, because it uses a
partial integration of dynamic equations. These integration leads to
appearance of arbitrary integration functions $g^{a}\left( \mathbf{\xi }%
\right) $. The wave function is constructed of these integration functions 
\cite{R1999}.

One can explain the situation as follows. It is well known, that the Schr%
\"{o}dinger equation can be written in the hydrodynamic form of
Madelung-Bohm \cite{M26,B52}. The wave function $\psi $ is presented in the
form 
\begin{equation}
\psi =\sqrt{\rho }\exp \left( i\varphi /\hbar \right)  \label{a2.1}
\end{equation}%
Substituting (\ref{a2.1}) in the Schr\"{o}dinger equation (\ref{a1.12}), one
obtains two real equations for dynamical variables $\rho $ and $\varphi $.
Taking gradient from the equation for $\varphi $ and introducing designation 
\begin{equation}
\mathbf{v=-}\frac{\hbar }{m}\mathbf{\nabla }\varphi ,\qquad \func{curl}%
\mathbf{v}=0  \label{a2.2}
\end{equation}%
one obtains four equations of the hydrodynamic type 
\begin{equation}
\frac{\partial \rho }{\partial t}+\mathbf{\nabla }\left( \rho \mathbf{v}%
\right) =0,\qquad \frac{d\mathbf{v}}{dt}\equiv \frac{\partial \mathbf{v}}{%
\partial t}+\left( \mathbf{v\nabla }\right) \mathbf{v}=-\frac{1}{m}\mathbf{%
\nabla }U_{\mathrm{B}}  \label{d1.14}
\end{equation}%
where $U_{\mathrm{B}}$ is the Bohm potential, defined by the relation 
\begin{equation}
U_{\mathrm{B}}=U\left( \rho ,\mathbf{\nabla }\rho ,\mathbf{\nabla }^{2}\rho
\right) =\frac{\hbar ^{2}}{8m\rho }\left( \frac{\left( \mathbf{\nabla }\rho
\right) ^{2}}{\rho }-2\mathbf{\nabla }^{2}\rho \right) =\mathbf{-}\frac{%
\hbar ^{2}}{2m\sqrt{\rho }}\mathbf{\nabla }^{2}\sqrt{\rho }  \label{d4.3}
\end{equation}%
Hydrodynamic equations (\ref{d1.14}) can be easily obtained from equations (%
\ref{d1.7}), (\ref{d1.8}). To obtain representation of equations (\ref{d1.14}%
), (\ref{d4.3}) in terms of wave function, one needs to integrate these
equations, because they have been obtained by means of differentiation of
the Schr\"{o}dinger equation. This integration can be easily produced, if
the condition (\ref{a2.2}) takes place and the fluid flow is non-rotational.

In the general case of vortical flow the integration is more complicated.
Nevertheless this integration has been produced \cite{R1999}, and one
obtains a more complicated equation (\ref{d6.14}), where two last terms
describe vorticity of the flow. The Schr\"{o}dinger equation (\ref{a1.12})
is a special case of the more general equation (\ref{d6.14}).

Note that the equation (\ref{d6.14}) is not linear, although it is invariant
with respect to transformation 
\begin{equation}
\psi \rightarrow \tilde{\psi}=A\psi ,\qquad A=\text{const}  \label{a2.3}
\end{equation}%
which admits one to normalize the wave function to any nonnegative quantity.
This property describes independence of the statistical ensemble on the
number of its constituents.

Representation of quantum mechanics as a statistical description of
classical indeterministic particles admits one to interpret all quantum
relations in terms of statistical description. This interpretation
distinguishes in some clauses from conventional (Copenhagen) interpretation
of quantum mechanics.

In any statistical description there are two different kinds of measurement,
which have different properties. Massive measurement (M-measurement) is
produced over all constituents of the statistical ensemble. A result of
M-measurement of the quantity $R$ is a distribution of the quantity $R$,
which can be predicted as a result of solution of dynamic equations for the
statistical ensemble.

Single measurement (S-measurement) is produced over one of constituents of
the statistical ensemble. A result of S-measurement of the quantity $R$ is
some random value of the quantity $R$, which cannot be predicted by the
theory. In the Copenhagen interpretation of the quantum mechanics the wave
function is supposed to describe a single particle (but not a statistical
ensemble of particles). As a result there is only one type of measurement,
which is considered sometimes as a M-measurement and sometimes as a
S-measurement. As far as M-measurement and S-measurement have different
properties, such an identification is a source of numerous contradictions
and paradoxes \cite{R2006b}.

Representation of quantum mechanics as a statistical description of the
indeterministic particles motion has two important consequences: (1)
elimination of quantum principles as laws of nature, (2) problem of
primordial stochastic motion of free particles.

\section{Deformation principle}

The idea, that a geometry is described completely by means of a distance
function (or world function) is very old. At first it was a metric space,
described by metric (distance). The metric has been restricted by a set of
conditions such as the triangle axiom and nonnegativity of the metric.
Condition of nonnegativity of metric does not permit to apply the metric
space for description of the space-time. The main defect of the metric
geometry and the distance geometry \cite{M28,B53} is impossibility of
construction of geometrical objects in terms of the world function or in
terms of the metric. Construction of geometrical objects in terms of the
world function is to be possible, because it is supposed that the geometry
is described completely by the world function and in terms of the world
function. Furthermore, a physical geometry is to admit a coordinateless
description.

Such a situation is possible, if one defines concepts of a geometry and
those of a geometrical objects correctly.

\textit{Definition 4.1}\emph{: }The physical geometry $\mathcal{G}=\left\{
\sigma ,\Omega \right\} $ is a point set $\Omega $ with the single-valued
function $\sigma $ \ on it 
\begin{equation}
\sigma :\qquad \Omega \times \Omega \rightarrow \mathbb{R},\qquad \sigma
\left( P,P\right) =0,\qquad \sigma \left( P,Q\right) =\sigma \left(
Q,P\right) ,\qquad P,Q\in \Omega  \label{b4.1}
\end{equation}

\textit{Definition 4.2:} Two physical geometries $\mathcal{G}_{1}=\left\{
\sigma _{1},\Omega _{1}\right\} $ \ and $\mathcal{G}_{2}=\left\{ \sigma
_{2},\Omega _{2}\right\} $ are equivalent $\left( \mathcal{G}_{1}\mathrm{eqv}%
\mathcal{G}_{2}\right) $ if the point set $\Omega _{1}\subseteq \Omega
_{2}\wedge \sigma _{1}=\sigma _{2}$, or $\Omega _{2}\subseteq \Omega
_{1}\wedge \sigma _{2}=\sigma _{1}$.

\textit{Remark:}\emph{\ }\ Coincidence of \ point sets $\Omega _{1}$ and $%
\Omega _{2}$ is not necessary for equivalence of geometries $\mathcal{G}_{1}$%
\textrm{\ }and\textrm{\ }$\mathcal{G}_{2}$. \ If one demands coincidence of $%
\Omega _{1}$ and $\Omega _{2}$ in the case equivalence of $\mathcal{G}_{1}$
and \ $\mathcal{G}_{2}$, then an elimination of one point $P$ from the point
set $\Omega _{1}$ turns the geometry $\mathcal{G}_{1}=\left\{ \sigma
_{1},\Omega _{1}\right\} $ into geometry $\mathcal{G}_{2}=\left\{ \sigma
_{1},\Omega _{1}\backslash P\right\} $, which appears to be not equivalent
to the geometry $\mathcal{G}_{1}$. Such a situation seems to be
inadmissible, because a geometry on a part $\omega \subset \Omega _{1}$ of
the point set $\Omega _{1}$ appears to be not equivalent to the geometry on
the whole point set $\Omega _{1}$.

According to definition \ the geometries $\mathcal{G}_{1}\mathcal{=}\left\{
\sigma ,\omega _{1}\right\} $ and $\mathcal{G}_{2}\mathcal{=}\left\{ \sigma
,\omega _{2}\right\} $ on parts\ of $\Omega $, $\omega _{1}\subset \Omega $
and $\omega _{2}\subset \Omega $ \ \ are equivalent $\left( \mathcal{G}_{1}%
\mathrm{eqv}\mathcal{G}\right) ,\left( \mathcal{G}_{2}\mathrm{eqv}\mathcal{G}%
\right) $ to the geometry $\mathcal{G}$, whereas the geometries $\mathcal{G}%
_{1}\mathcal{=}\left\{ \sigma ,\omega _{1}\right\} $ and $\mathcal{G}_{2}%
\mathcal{=}\left\{ \sigma ,\omega _{2}\right\} $ are not equivalent, in
general, if $\omega _{1}\nsubseteqq \omega _{2}$ and $\omega _{2}\nsubseteqq
\omega _{1}$. Thus, the relation of equivalence is intransitive, in general.
The space-time geometry may vary in different regions of the space-time. It
means, that a physical body, described as a geometrical object, may evolve
in such a way, that it appears in regions with different space-time geometry.

\textit{Definition 4.3:}\emph{\ }A geometrical object $g_{\mathcal{P}_{n}}$
of the geometry $\mathcal{G=}\left\{ \sigma ,\Omega \right\} $ is a subset $%
g_{\mathcal{P}_{n}}\subset \Omega $ of the point set $\Omega $. This
geometrical object $g_{\mathcal{P}_{n}}$ is a set of roots $R\in \Omega $ of
the function $F_{\mathcal{P}_{n}}$%
\begin{equation*}
F_{\mathcal{P}_{n}}:\qquad \Omega \rightarrow \mathbb{R}
\end{equation*}%
where%
\begin{eqnarray}
F_{\mathcal{P}_{n}} &:&\quad F_{\mathcal{P}_{n}}\left( R\right) =G_{\mathcal{%
P}_{n}}\left( u_{1},u_{2},...u_{s}\right) ,\qquad s=\frac{1}{2}\left(
n+1\right) \left( n+2\right)  \label{b4.3} \\
u_{l} &=&\sigma \left( w_{i},w_{k}\right) ,\qquad i,k=0,1,...n+1,\qquad
l=1,2,...\frac{1}{2}\left( n+1\right) \left( n+2\right)  \label{b4.4} \\
w_{k} &=&P_{k}\in \Omega ,\qquad k=0,1,...n,\qquad w_{n+1}=R\in \Omega
\label{b4.4a}
\end{eqnarray}%
Here $\mathcal{P}_{n}=\left\{ P_{0},P_{1},...,P_{n}\right\} \subset \Omega $
are \ $n+1$ points which are parameters, determining the geometrical object $%
g_{\mathcal{P}_{n}}$%
\begin{equation}
g_{\mathcal{P}_{n}}=\left\{ R|F_{\mathcal{P}_{n}}\left( R\right) =0\right\}
,\qquad R\in \Omega ,\qquad \mathcal{P}_{n}\in \Omega ^{n+1}  \label{b4.2}
\end{equation}%
$F_{\mathcal{P}_{n}}\left( R\right) =G_{\mathcal{P}_{n}}\left(
u_{1},u_{2},...u_{s}\right) $ is an arbitrary function of $\frac{1}{2}\left(
n+1\right) \left( n+2\right) $ arguments \ $u_{s}$ and of \ $n+1$ parameters 
$\mathcal{P}_{n}$. The set $\mathcal{P}_{n}$ of the geometric object
parameters will be referred to as the skeleton of the geometrical object.
The subset $g_{\mathcal{P}_{n}}$ will be referred to as the envelope of the
skeleton. One skeleton may have many envelopes. When a particle is
considered as a geometrical object, its motion in the space-time is
described mainly by the skeleton $\mathcal{P}_{n}$. The shape of the
envelope is of no importance in the first approximation.

\textit{Remark: }Arbitrary subset of the point set $\Omega $ is not a
geometrical object, in general. It is supposed, that physical bodies may
have a shape of a geometrical object only, because only in this case one can
identify identical physical bodies (geometrical objects) in different
space-time geometries.

Existence of the same geometrical objects in different space-time regions,
having different geometries, arises the question on equivalence of
geometrical objects in different space-time geometries. Such a question was
not arisen before, because one does not consider such a situation, when the
physical body moves from one space-time region to another space-time region,
having another space-time geometry. In general, mathematical technique of
the conventional space-time geometry is not applicable for simultaneous
consideration of several different geometries of different space-time
regions.

We can perceive the space-time geometry only via motion of physical bodies
in the space-time, or via construction of geometrical objects corresponding
to these physical bodies. As it follows from the \textit{definition 4.3} of
the geometrical object, the function \ $F$ as a function of its arguments
(of world functions of different points) is the same in all physical
geometries. It means, that a geometrical object $\mathcal{O}_{1}$ in the
geometry $\mathcal{G}_{1}=\left\{ \sigma _{1},\Omega _{1}\right\} $ is
obtained from the same geometrical object $\mathcal{O}_{2}$ in the geometry $%
\mathcal{G}_{2}=\left\{ \sigma _{2},\Omega _{2}\right\} $ by means of the
replacement $\sigma _{2}\rightarrow \sigma _{1}$ in the definition of this
geometrical object.

As far as the physical geometry is determined by its geometrical objects
construction, a physical geometry $\mathcal{G}=\left\{ \sigma ,\Omega
\right\} $ can be obtained from some known standard geometry $\mathcal{G}_{%
\mathrm{st}}=\left\{ \sigma _{\mathrm{st}},\Omega \right\} $ by means a
deformation of the standard geometry $\mathcal{G}_{\mathrm{st}}$.
Deformation of the standard geometry $\mathcal{G}_{\mathrm{st}}$ is realized
by the replacement $\sigma _{\mathrm{st}}\rightarrow \sigma $ in all
definitions of the geometrical objects in the standard geometry. The proper
Euclidean geometry is an axiomatizable geometry. It has been constructed by
means of the Euclidean method as a logical construction. The proper
Euclidean geometry is a physical geometry. It may be used as a standard
geometry $\mathcal{G}_{\mathrm{st}}$. Construction of a physical geometry as
a deformation of the proper Euclidean geometry will be referred to as the
deformation principle. The most physical geometries are nonaxiomatizable
geometries. They can be constructed only by means of the deformation
principle.

Description of the elementary particle motion in the space-time contains
only the particle skeleton $\mathcal{P}_{n}=\left\{
P_{0},P_{1},...P_{n}\right\} $. The form of the function (\ref{b4.3}) is of
no importance in the first approximation. In the elementary particle
dynamics only equivalence of vectors $\mathbf{P}_{i}\mathbf{P}_{k}$, \ $%
i,k=0,1,...n$ is essential. These vectors are defined by the particle
skeleton $\mathcal{P}_{n}$.

The equivalence $\left( \mathbf{P}_{0}\mathbf{P}_{1}\mathrm{eqv}\mathbf{Q}%
_{0}\mathbf{Q}_{1}\right) $ of two vectors $\mathbf{P}_{0}\mathbf{P}_{1}$
and $\mathbf{Q}_{0}\mathbf{Q}_{1}$ is defined by the relations%
\begin{equation}
\left( \mathbf{P}_{0}\mathbf{P}_{1}\mathrm{eqv}\mathbf{Q}_{0}\mathbf{Q}%
_{1}\right) :\qquad \left( \mathbf{P}_{0}\mathbf{P}_{1}\mathrm{.}\mathbf{Q}%
_{0}\mathbf{Q}_{1}\right) =\left\vert \mathbf{P}_{0}\mathbf{P}%
_{1}\right\vert \cdot \left\vert \mathbf{Q}_{0}\mathbf{Q}_{1}\right\vert
\wedge \left\vert \mathbf{P}_{0}\mathbf{P}_{1}\right\vert =\left\vert 
\mathbf{Q}_{0}\mathbf{Q}_{1}\right\vert  \label{b4.5}
\end{equation}%
where%
\begin{equation}
\left\vert \mathbf{P}_{0}\mathbf{P}_{1}\right\vert =\sqrt{2\sigma \left(
P_{0},P_{1}\right) }  \label{b4.6}
\end{equation}%
and the scalar product $\left( \mathbf{P}_{0}\mathbf{P}_{1}\mathrm{.}\mathbf{%
Q}_{0}\mathbf{Q}_{1}\right) $ is defined by the relation (\ref{b1.9})%
\begin{equation}
\left( \mathbf{P}_{0}\mathbf{P}_{1}.\mathbf{Q}_{0}\mathbf{Q}_{l}\right)
=\sigma \left( P_{0},Q_{l}\right) +\sigma \left( P_{0},Q_{l}\right) -\sigma
\left( P_{0},Q_{l}\right) -\sigma \left( P_{0},Q_{l}\right)  \label{b4.7}
\end{equation}

Skeletons $\mathcal{P}_{n}=\left\{ P_{0},P_{1},...P_{n}\right\} $ and $%
\mathcal{P}_{n}^{\prime }=\left\{ P_{0}^{\prime },P_{1}^{\prime
},...P_{n}^{\prime }\right\} $ may belong to the same geometrical object, if 
\begin{equation}
\left\vert \mathbf{P}_{i}\mathbf{P}_{k}\right\vert =\left\vert \mathbf{P}%
_{i}^{\prime }\mathbf{P}_{k}^{\prime }\right\vert ,\qquad i,k=0,1,...n
\label{b4.9}
\end{equation}%
i.e. lengths of all vectors $\mathbf{P}_{i}\mathbf{P}_{k}$ and $\mathbf{P}%
_{i}^{\prime }\mathbf{P}_{k}^{\prime }$ are equal. However, it is not
sufficient for equivalence of skeletons $\mathcal{P}_{n}$ and $\mathcal{P}%
_{n}^{\prime }$ .

Skeletons $\mathcal{P}_{n}=\left\{ P_{0},P_{1},...P_{n}\right\} $ and $%
\mathcal{P}_{n}^{\prime }=\left\{ P_{0}^{\prime },P_{1}^{\prime
},...P_{n}^{\prime }\right\} $ are equivalent 
\begin{equation}
\left( \mathcal{P}_{n}\mathrm{eqv}\mathcal{P}_{n}^{\prime }\right) :\qquad 
\text{if}\ \ \ \ \mathbf{P}_{i}\mathbf{P}_{k}\mathrm{eqv}\mathbf{P}%
_{i}^{\prime }\mathbf{P}_{k}^{\prime },\qquad i,k=0,1,,...n  \label{b.4.10}
\end{equation}%
In other words, the equality of skeletons needs equality of the lengths of
vectors $\mathbf{P}_{i}\mathbf{P}_{k}$ and $\mathbf{P}_{i}^{\prime }\mathbf{P%
}_{k}^{\prime }$ and equality of their mutual orientations.

\section{Multivariance}

The physical geometry has the property, called multivariance. It means that
at the point $P_{0}$ there are many vectors $\mathbf{P}_{0}\mathbf{P}_{1}$, $%
\mathbf{P}_{0}\mathbf{P}_{1}^{\prime }$, $\mathbf{P}_{0}\mathbf{P}%
_{1}^{\prime \prime },...$which are equivalent to the vector $\mathbf{Q}_{0}%
\mathbf{Q}_{1}$ at the point $Q_{0}$, but they are not equivalent between
themselves. The proper Euclidean geometry has not the property of
multivariance. In the proper Euclidean geometry there is only one vector $%
\mathbf{P}_{0}\mathbf{P}_{1}$ at the point $P_{0}$, which is equivalent to
the vector $\mathbf{Q}_{0}\mathbf{Q}_{1}$ at the point $Q_{0}$.

Multivariance is connected formally with the definition of the vector
equivalence via algebraic relations (\ref{b4.5}) - (\ref{b4.7}). If vector $%
\mathbf{Q}_{0}\mathbf{Q}_{1}$ is given, and it is necessary to determine the
equivalent vector $\mathbf{P}_{0}\mathbf{P}_{1}$ at the point $P_{0}$, one
needs to solve two equations (\ref{b4.5}) with respect to the point $P_{1}$.
If the two equations have a unique solution, one has only one equivalent
vector $\mathbf{P}_{0}\mathbf{P}_{1}$ (single-variance). If there are many
solutions, one has many vectors $\mathbf{P}_{0}\mathbf{P}_{1}$, $\mathbf{P}%
_{0}\mathbf{P}_{1}^{\prime }$, $\mathbf{P}_{0}\mathbf{P}_{1}^{\prime \prime
},..$, which are equivalent to vector $\mathbf{Q}_{0}\mathbf{Q}_{1}$
(multivariance). It is possible such a case, when there are no solutions. In
this case one has zero-variance.

Multivariance of the space-time geometry leads to splitting of one world
chain into many stochastic world chains. As a result the multivariance of
the space-time geometry in microcosm leads to appearance of quantum effects.

Zero-variance appears in the case of many-point skeletons. It is interesting
in that relation, that it may forbid existence of elementary particles with
many-point skeletons.

\section{Discreteness of the space-time geometry}

The world function (\ref{b1.3}) describes a completely discrete geometry.
However, the space-time geometry may discrete only partly. In the discrete
geometry one may introduce the point density $\rho =d\sigma _{\mathrm{M}%
}/d\sigma _{\mathrm{d}}$ with respect to point density in the geometry of
Minkowski. The discrete geometry may be described by the relative points
density%
\begin{equation}
\rho \left( \sigma _{\mathrm{d}}\right) =\frac{d\sigma _{\mathrm{M}}\left(
\sigma _{\mathrm{d}}\right) }{d\sigma _{\mathrm{d}}}=\left\{ 
\begin{array}{ccc}
0 & \text{if} & 0<\left\vert \sigma _{\mathrm{d}}\right\vert <\frac{\lambda
_{0}^{2}}{2} \\ 
1 & \text{if} & \left\vert \sigma _{\mathrm{M}}\right\vert \geq \frac{%
\lambda _{0}^{2}}{2}%
\end{array}%
\right.  \label{b6.1}
\end{equation}%
For close points the relative point density of the discrete geometry
vanishes, and this circumstance is considered as a discreteness of the
geometry. However, the discreteness may not be complete.

Let us consider the space-time geometry with the world function $\sigma _{%
\mathrm{g}}$%
\begin{equation}
\sigma _{\mathrm{g}}=\sigma _{\mathrm{M}}+\frac{\lambda _{0}^{2}}{2}\left\{ 
\begin{array}{ccc}
\text{sgn}\left( \sigma _{\mathrm{M}}\right) & \text{if} & \left\vert \sigma
_{\mathrm{M}}\right\vert \geq \sigma _{0} \\ 
\frac{\sigma _{\mathrm{M}}}{\sigma _{0}} & \text{if} & \left\vert \sigma _{%
\mathrm{M}}\right\vert <\sigma _{0}%
\end{array}%
\right. ,\quad \lambda _{0},\sigma _{0}=\text{const}  \label{b6.2}
\end{equation}%
The relative point density in the geometry (\ref{b6.2}) has the form%
\begin{equation}
\rho \left( \sigma _{\mathrm{g}}\right) =\frac{d\sigma _{\mathrm{M}}\left(
\sigma _{\mathrm{g}}\right) }{d\sigma _{\mathrm{g}}}=\left\{ 
\begin{array}{ccc}
1 & \text{if} & \left\vert \sigma _{\mathrm{g}}\right\vert \geq \sigma _{0}+%
\frac{\lambda _{0}^{2}}{2} \\ 
\frac{\sigma _{0}}{\sigma _{0}+\frac{\lambda _{0}^{2}}{2}} & \text{if} & 
\left\vert \sigma _{\mathrm{g}}\right\vert <\sigma _{0}+\frac{\lambda
_{0}^{2}}{2}%
\end{array}%
\right.  \label{b6.3}
\end{equation}%
If $\sigma _{0}\ll \lambda _{0}^{2}$ the relative point density in the
region, where $\left\vert \sigma _{\mathrm{g}}\right\vert \in \left(
0,\sigma _{0}+\frac{\lambda _{0}^{2}}{2}\right) $ is much less, than $1$. If 
$\sigma _{0}\rightarrow 0$, the relative point density (\ref{b6.3}) tends to
(\ref{b6.1}). The geometry (\ref{b6.2}) should be qualified as a partly
discrete space-time geometry. We shall refer to the geometry (\ref{b6.2}) as
a granular geometry. In the granular space-time geometry the relative
density of points, separated by small distance (less, than $\lambda _{0}$),
is much less than the relative density of other points. The granular
geometry, described by the world function 
\begin{eqnarray}
\sigma _{\mathrm{g}} &=&\sigma _{\mathrm{M}}+\frac{\lambda _{0}^{2}}{2}%
\left\{ 
\begin{array}{ccc}
\text{sgn}\left( \sigma _{\mathrm{M}}\right) & \text{if} & \left\vert \sigma
_{\mathrm{M}}\right\vert >\sigma _{0} \\ 
f\left( \frac{\sigma _{\mathrm{M}}}{\sigma _{0}}\right) & \text{if} & 
\left\vert \sigma _{\mathrm{M}}\right\vert \leq \sigma _{0}%
\end{array}%
\right. ,\quad \lambda _{0},\sigma _{0}=\text{const}  \label{b6.4} \\
f\left( x\right) &=&-f\left( -x\right) ,\quad f\left( 1\right) =1  \notag
\end{eqnarray}%
is a generalization of the the geometry (\ref{b6.2}).

\section{Elementary particle dynamics}

Dynamics of elementary particles in the granular space-time geometry is
considered in \cite{R2008b}. The state of an elementary particle is
described by its skeleton $\mathcal{P}_{n}=\left\{
P_{0},P_{1},...P_{n}\right\} $, consisting of $n+1$ space-time points. Such
description of the particle state is complete in the sense, that it does not
need parameters of the particle (mass, charge, spin, etc.). All this
information is described by the disposition of points in the skeleton. It
means a geometrization of parameters of the elementary particles. Besides,
the conventional description of the particle state as a point in the phase
space is nonrelativistic. The granular geometry is multivariant, in general.
The particle motion is stochastic, and the limit (\ref{b1.7}), which
determines the particle momentum, does not exist. Thus, \textit{to satisfy
the relativity principles, we are forced to describe the particle state by
its skeleton}.

Evolution of the particle state is described by the world chain $\mathcal{C}$%
, consisting of connected skeletons $\mathcal{P}_{n}^{\left( s\right)
}=\left\{ P_{0}^{\left( s\right) },P_{1}^{\left( s\right) },...P_{n}^{\left(
s\right) }\right\} $, $s=...-1,0,1,...$%
\begin{equation}
\mathcal{C=}\dbigcup\limits_{s}\mathcal{P}_{n}^{\left( s\right) },\quad
P_{1}^{\left( s\right) }=P_{0}^{\left( s+1\right) },\quad s=...-1,0,1,...
\label{b7.1}
\end{equation}%
Connection between skeletons of the world chain arises, because the second
point $P_{1}^{\left( s\right) }$ of the $\mathit{s}$th skeleton coincides
with the first point $P_{0}^{\left( s+1\right) }$ of the $\left( s+1\right) $%
th skeleton. The vector $\mathbf{P}_{0}^{\left( s\right) }\mathbf{P}%
_{1}^{\left( s\right) }=\mathbf{P}_{0}^{\left( s\right) }\mathbf{P}%
_{0}^{\left( s+1\right) }$ will be referred to as the leading vector,
determining the shape of the world chain. All skeletons of the chain are
similar in the sense, that 
\begin{equation}
\left\vert \mathbf{P}_{i}^{\left( s\right) }\mathbf{P}_{k}^{\left( s\right)
}\right\vert =\mu _{ik}=\text{const, }\quad i,k=0,1,..n,\quad s=...-1,0,1,...
\label{b7.2}
\end{equation}

\textit{Definition:} Two vectors $\mathbf{P}_{0}\mathbf{P}_{1}$ and $\mathbf{%
Q}_{0}\mathbf{Q}_{1}$ are equivalent $\left( \mathbf{P}_{0}\mathbf{P}_{1}%
\mathrm{eqv}\mathbf{Q}_{0}\mathbf{Q}_{1}\right) $, if%
\begin{equation}
\left( \mathbf{P}_{0}\mathbf{P}_{1}.\mathbf{Q}_{0}\mathbf{Q}_{1}\right)
=\left\vert \mathbf{P}_{0}\mathbf{P}_{1}\right\vert \cdot \left\vert \mathbf{%
Q}_{0}\mathbf{Q}_{1}\right\vert \wedge \left\vert \mathbf{P}_{0}\mathbf{P}%
_{1}\right\vert =\left\vert \mathbf{Q}_{0}\mathbf{Q}_{1}\right\vert
\label{b7.2a}
\end{equation}%
If the particle is free, then the skeleton motion is progressive (i.e.
motion without rotation), and orientation of adjacent skeletons $\mathcal{P}%
_{n}^{\left( s\right) }$, $\mathcal{P}_{n}^{\left( s+1\right) }$ is the same.%
\begin{eqnarray}
\left( \mathbf{P}_{i}^{\left( s\right) }\mathbf{P}_{k}^{\left( s\right) }.%
\mathbf{P}_{i}^{\left( s+1\right) }\mathbf{P}_{k}^{\left( s+1\right)
}\right) &=&\left\vert \mathbf{P}_{i}^{\left( s\right) }\mathbf{P}%
_{k}^{\left( s\right) }\right\vert \cdot \left\vert \mathbf{P}_{i}^{\left(
s+1\right) }\mathbf{P}_{k}^{\left( s+1\right) }\right\vert =\mu _{ik}^{2},
\label{b7.3} \\
i,k &=&0,1,..n,\quad s=...-1,0,1,...  \notag
\end{eqnarray}%
Equations (\ref{b7.2}), (\ref{b7.3}) means that the adjacent skeletons of
the world chain are equivalent $\mathcal{P}_{n}^{\left( s\right) }$eqv$%
\mathcal{P}_{n}^{\left( s+1\right) }$, $s=...-1,0,1,...$The adjacent
skeletons are equivalent, if corresponding vectors of adjacent skeletons are
equivalent%
\begin{equation}
\left( \mathbf{P}_{i}^{\left( s\right) }\mathbf{P}_{k}^{\left( s\right) }%
\text{eqv}\mathbf{P}_{i}^{\left( s+1\right) }\mathbf{P}_{k}^{\left(
s+1\right) }\right) ,\quad i,k=0,1,..n,\quad s=...-1,0,1,..  \label{b7.4}
\end{equation}%
One obtains $n\left( n+1\right) $ difference dynamic equations (\ref{b7.4})
(or (\ref{b7.2}), (\ref{b7.3})), which describe evolution of the particle
state. Introducing a coordinate system, one obtains $nD$ dynamic variables,
whose values are to be determined by dynamic equations (\ref{b7.4}). Here $D$
is the dimension of the space-time (the number of coordinates, describing
the point position). In particular, in the case of a pointlike particle,
whose state is described by two points $P_{0}$, $P_{1}$, the number of
dynamic equations $n_{\mathrm{d}}=2$, whereas in the 4D-space-time the
number of variable $n_{\mathrm{v}}=4$. In the multivariant space-time the
dynamic equations have many solutions. As a result the world chain appears
to be multivariant (stochastic).

In the Riemannian space-time and in the space-time of Minkowski the world
chain can be approximated by a world line, provided characteristic lengths
of the problem are much larger, than the lengths of the world chain links.
In this case the dynamic equations (\ref{b7.4}) are reduced to ordinary
differential equations. If the world line is timelike \cite{R2008b}, the
solution of dynamic equations appears to be unique. If the vectors $\mathbf{P%
}_{0}^{\left( s\right) }\mathbf{P}_{1}^{\left( s\right) }$ are spacelike,
dynamic equations have many solutions even in the Riemannian space-time. It
is connected with the circumstance, that the Riemannian geometry as well as
the geometry of Minkowski is multivariant with respect to spacelike vectors.
At the conventional approach the spacelike world lines are not considered at
all. Such world lines are inadmissible by definition (It is a postulate).

One attempted to obtain differential dynamic equations for a pointlike
particle in \cite{R2010b}. At first one obtained equation for free pointlike
particle in the space-time of Minkowski. It is only one equation, whereas in
the conventional approach one has three equations for the velocity
components $\mathbf{\beta }=\mathbf{v}/c.$ This equation has the form%
\begin{equation}
\mathbf{\dot{\beta}}^{2}+\frac{\left( \mathbf{\beta \dot{\beta}}\right) ^{2}%
}{1-\mathbf{\beta }^{2}}=0,\quad \mathbf{\dot{\beta}\equiv }\frac{d\mathbf{%
\beta }}{dt}  \label{b7.5}
\end{equation}

Let us introduce designation%
\begin{equation}
\mathbf{\beta \dot{\beta}=}\sqrt{\mathbf{\beta }^{2}\mathbf{\dot{\beta}}^{2}}%
\cos \phi  \label{b7.6}
\end{equation}%
where $\phi $ is the angle between vectors $\mathbf{\beta }$ and $\mathbf{%
\dot{\beta}}$\textbf{.} The equation (\ref{b7.5}) takes the form 
\begin{equation}
\mathbf{\dot{\beta}}^{2}\left( 1+\frac{\mathbf{\beta }^{2}\cos ^{2}\phi }{1-%
\mathbf{\beta }^{2}}\right) =0  \label{b7.7}
\end{equation}

If the world line is timelike $\mathbf{\beta }^{2}<1$ , and cos$^{2}\phi
\leq 1$, then the bracket in (\ref{b7.7}) is positive and one concludes from
(\ref{b7.7}), that 
\begin{equation}
\mathbf{\dot{\beta}}^{2}=0  \label{b7.8}
\end{equation}%
One obtains three equations from one equation (\ref{b7.8}) 
\begin{equation}
\mathbf{\dot{\beta}}\equiv c^{-1}\frac{d\mathbf{v}}{dt}=0  \label{b7.9}
\end{equation}

If the world line is spacelike, then $\mathbf{\beta }^{2}>1$, and the
bracket in (\ref{b7.8}) vanishes at 
\begin{equation}
\cos ^{2}\phi =\frac{\mathbf{\beta }^{2}-1}{\mathbf{\beta }^{2}}<1
\label{b7.10}
\end{equation}%
The acceleration $\mathbf{\dot{v}}=c\mathbf{\dot{\beta}}$ becomes indefinite
at this value of the angle $\phi $ between $\mathbf{\dot{\beta}}$ and $%
\mathbf{\beta }$. It should be interpreted as impossibility of spacelike$%
\mathbf{\ }$world lines. At the conventional approach such an impossibility
of spacelike world lines is simply postulated.

Such a result is rather evident, because the space-time of Minkowski is
single-variant with respect to timelike vectors and it is multivariant with
respect to spacelike vectors. For timelike vectors one can obtain three
dynamic equations (\ref{b7.9}) from one equation (\ref{b7.7}). For spacelike
particles it is impossible.

Another example is considered in the paper \cite{R2010b}. Motion of
pointlike particle in the gravitational field of a massive sphere of the
mass $M$ is considered. In the Newtonian approximation the world function $%
\sigma \left( t,\mathbf{y;}t^{\prime },\mathbf{y}^{\prime }\right) $ between
the points with coordinates $\left( t,\mathbf{y}\right) $ and $\left(
t^{\prime },\mathbf{y}^{\prime }\right) $ has the form%
\begin{equation}
\sigma \left( t,\mathbf{y;}t^{\prime },\mathbf{y}^{\prime }\right) =\frac{1}{%
2}\left( c^{2}-\frac{2GM}{\sqrt{\mathbf{x}^{2}}}\right) \left( t-t^{\prime
}\right) ^{2}-\frac{1}{2}\left( \mathbf{y}-\mathbf{y}^{\prime }\right) ^{2}
\label{b7.11}
\end{equation}%
where $G$ is the gravitational constant, and 
\begin{equation}
\mathbf{x=}\frac{\mathbf{y+y}^{\prime }}{2}  \label{b7.12}
\end{equation}%
Metric tensor has the conventional form 
\begin{equation}
g_{ik}=g_{ik}\left( \mathbf{x}\right) =\text{diag}\left( c^{2}-\frac{2GM}{%
\sqrt{\mathbf{x}^{2}}},-1,-1,-1\right)  \label{b7.14}
\end{equation}%
but the space-time geometry, described by the world function (\ref{b7.11})
is non-Riemannian.

The Riemannian geometry is conceptually defective in the sense, the world
function of the Riemannian geometry with metric tensor (\ref{b7.14}) is
multivalued, whereas the world function is to be single-valued. But the
Riemannian geometry is single-variant with respect to timelike vectors,
having common origin. As a result the timelike world chains in the
Riemannian space-time geometry are deterministic. They can be replaced by
deterministic world lines.

The space-time geometry (\ref{b7.11}) is multivariant in general, but the
world function (\ref{b7.11}) is single-valued. The world function (\ref%
{b7.11}) is obtained in the extended general relativity, when one eliminates
the unfounded restriction that the space-time geometry is to be a Riemannian
geometry \cite{R2010c}.

To obtain differential dynamic equations for a free particle, one considers
two connected links of the world chain, defined by the points $%
P_{0},P_{1},P_{2}$, having coordinates%
\begin{equation}
P_{0}=\left\{ y-dy_{1}\right\} ,\qquad P_{1}=\left\{ y\right\} ,\qquad
P_{2}=\left\{ y+dy_{2}\right\}  \label{a3.1}
\end{equation}%
where 
\begin{equation}
y=\left\{ t,\mathbf{y}\right\} ,\qquad dy_{1}=\left\{ dt_{1},d\mathbf{y}%
_{1}\right\} ,\qquad dy_{2}=\left\{ dt_{2},d\mathbf{y}_{2}\right\}
\label{a3.2}
\end{equation}%
are coordinates in some inertial coordinate system. Dynamic equations (\ref%
{b7.2}), (\ref{b7.3}) have the form%
\begin{eqnarray}
\mathrm{\ }\sigma \left( y,y-dy_{1}\right) &=&\sigma \left( y,y+dy_{2}\right)
\label{a3.3} \\
\mathrm{\ }4\sigma \left( y,y-dy_{1}\right) &=&\sigma \left(
y-dy_{1},y+dy_{2}\right)  \label{a3.4}
\end{eqnarray}

Let us introduce designations 
\begin{eqnarray}
\mathbf{v}_{1} &=&\frac{d\mathbf{y}_{1}}{dt_{1}},\qquad \mathbf{v}_{2}=\frac{%
d\mathbf{y}_{2}}{dt_{2}},\qquad \mathbf{\beta }_{1}\mathbf{=}\frac{\mathbf{v}%
_{1}}{c},\qquad \mathbf{\beta }_{2}\mathbf{=}\frac{\mathbf{v}_{2}}{c}
\label{a3.7} \\
\mathbf{\beta }_{1} &=&\mathbf{\beta -}\frac{1}{2}\mathbf{\dot{\beta}}%
dt,\qquad \mathbf{\beta }_{2}=\mathbf{\beta +}\frac{1}{2}\mathbf{\dot{\beta}}%
dt,\qquad \mathbf{\dot{\beta}\equiv }\frac{d\mathbf{\beta }}{dt},\qquad dt=%
\frac{dt_{1}+dt_{2}}{2}  \label{a3.8}
\end{eqnarray}%
where 
\begin{equation}
\mathbf{v=}c\mathbf{\beta }\qquad \mathbf{\dot{v}}=c\mathbf{\dot{\beta}}
\label{a3.8a}
\end{equation}%
\begin{equation}
V=V\left( \mathbf{y}\right) =\frac{GM}{\sqrt{\left( \mathbf{y}\right) ^{2}}}%
,\qquad U=U\left( \mathbf{y}\right) =\frac{V\left( \mathbf{y}\right) }{c^{2}}
\label{a4.3}
\end{equation}
Using designations (\ref{a3.7}) - (\ref{a4.3}), transforming two equations (%
\ref{a3.3}), (\ref{a3.4}), and considering $dt,d\mathbf{y}_{1},d\mathbf{y}%
_{2}$ as infinitesimal quantities, one obtains after simplifications%
\begin{equation}
\frac{1}{2}\mathbf{\dot{\beta}}^{2}\left( dt\right) ^{2}-c\mathbf{\dot{\beta}%
\nabla }U\left( dt\right) ^{2}+\frac{1}{2}\frac{\left( c\mathbf{\beta \nabla 
}U+\mathbf{\beta \dot{\beta}}\right) ^{2}}{1-2U-\mathbf{\beta }^{2}}\left(
dt\right) ^{2}+\frac{c^{2}}{2}\beta ^{\alpha }\beta ^{\beta }\partial
_{\alpha }\partial _{\beta }U\left( dt\right) ^{2}=0  \label{a4.24}
\end{equation}%
where 
\begin{equation*}
\partial _{\alpha }\equiv \frac{\partial }{\partial y^{\alpha }}
\end{equation*}%
Note, that the terms of the order of $dt$ disappear.

In terms of variables $\mathbf{v,\dot{v},}V$, defined by relations (\ref%
{a3.8a}), (\ref{a4.3}) the relation (\ref{a4.24}) has the form%
\begin{equation}
\frac{1}{2}\mathbf{\dot{v}}^{2}-\mathbf{\dot{v}\nabla }V+\frac{1}{2}\frac{%
\left( \mathbf{v\nabla }V+\mathbf{v\dot{v}}\right) ^{2}}{c^{2}\left(
1-2c^{-2}V-c^{-2}\mathbf{v}^{2}\right) }+\frac{1}{2c^{2}}v^{\alpha }v^{\beta
}\partial _{\alpha }\partial _{\beta }V=0  \label{a4.25}
\end{equation}

One obtains in the nonrelativistic approximation 
\begin{equation}
\frac{1}{2}\mathbf{\dot{v}}^{2}-\mathbf{\dot{v}\nabla }V=0  \label{a4.26}
\end{equation}%
It is evident, that one cannot determine three components of vector $\mathbf{%
\dot{v}}$ from one equation (\ref{a4.26}). One can determine only mean value 
$\left\langle \mathbf{\dot{v}}\right\rangle $ of vector $\mathbf{\dot{v}}$%
\textbf{, }choosing some principle of averaging.

Let us represent $\mathbf{v}$ in the form 
\begin{equation}
\mathbf{\dot{v}}=\mathbf{\dot{v}}_{\parallel }+\mathbf{\dot{v}}_{\perp
},\qquad \mathbf{\dot{v}}_{\parallel }=\mathbf{\nabla }V\frac{\left( \mathbf{%
\dot{v}\nabla }V\right) }{\left\vert \mathbf{\nabla }V\right\vert ^{2}}%
,\qquad \mathbf{\dot{v}}_{\perp }=\mathbf{\dot{v}-\nabla }V\frac{\left( 
\mathbf{\dot{v}\nabla }V\right) }{\left\vert \mathbf{\nabla }V\right\vert
^{2}}  \label{a4.27}
\end{equation}%
where $\mathbf{v}_{\parallel }$ and $\mathbf{v}_{\perp }$ are components of $%
\mathbf{v}$, which are parallel to $\mathbf{\nabla }V$ and perpendicular to $%
\mathbf{\nabla }V$ correspondingly. It follows from (\ref{a4.26}) 
\begin{equation}
\mathbf{\dot{v}}_{\parallel }^{2}-2\mathbf{\dot{v}}_{\parallel }\mathbf{%
\nabla }V+\mathbf{\dot{v}}_{\perp }^{2}=0  \label{a4.28}
\end{equation}%
Let 
\begin{equation*}
\dot{v}_{\parallel }=\frac{\mathbf{\dot{v}\nabla }V}{\left\vert \mathbf{%
\nabla }V\right\vert }=\frac{\mathbf{\dot{v}}_{\parallel }\mathbf{\nabla }V}{%
\left\vert \mathbf{\nabla }V\right\vert },\qquad \mathbf{\dot{v}}_{\parallel
}=\mathbf{\nabla }V\frac{\left( \mathbf{\dot{v}\nabla }V\right) }{\left\vert 
\mathbf{\nabla }V\right\vert ^{2}}=\frac{\mathbf{\nabla }V}{\left\vert 
\mathbf{\nabla }V\right\vert }\dot{v}_{\parallel }
\end{equation*}%
The equation (\ref{a4.28}) may be rewritten in the form 
\begin{equation}
\dot{v}_{\parallel }^{2}-2\dot{v}_{\parallel }\left\vert \mathbf{\nabla }%
V\right\vert +\mathbf{\dot{v}}_{\perp }^{2}=0  \label{b7.15}
\end{equation}%
or%
\begin{equation}
\dot{v}_{\parallel }=\left\vert \mathbf{\nabla }V\right\vert \pm \sqrt{%
\left\vert \mathbf{\nabla }V\right\vert ^{2}-\mathbf{\dot{v}}_{\perp }^{2}}
\label{a4.29}
\end{equation}%
It follows from (\ref{a4.29}), that%
\begin{equation}
0<\mathbf{\dot{v}}_{\perp }^{2}\leq \left\vert \mathbf{\nabla }V\right\vert
^{2},\qquad 0<\dot{v}_{\parallel }<2\left\vert \mathbf{\nabla }V\right\vert
\label{a4.30}
\end{equation}%
The quantity $\dot{v}_{\parallel }$ vibrates around its mean value $%
\left\langle \dot{v}_{\parallel }\right\rangle =\left\vert \mathbf{\nabla }%
V\right\vert $. 
\begin{equation}
\left\langle \mathbf{\dot{v}}_{\parallel }\right\rangle =\frac{\mathbf{%
\nabla }V}{\left\vert \mathbf{\nabla }V\right\vert }\left\langle \dot{v}%
_{\parallel }\right\rangle =\mathbf{\nabla }V  \label{b7.16}
\end{equation}%
Taking into account a symmetry and supposing that $\left\langle \mathbf{\dot{%
v}}_{\perp }\right\rangle =0$, one obtains, that%
\begin{equation}
\left\langle \mathbf{\dot{v}}\right\rangle =\left\langle \mathbf{\dot{v}}%
_{\parallel }\right\rangle =\mathbf{\nabla }V=\mathbf{\nabla }\frac{GM}{r}%
,\quad r=\left\vert \mathbf{y}\right\vert  \label{a4.31}
\end{equation}

In the general case one obtains instead of (\ref{a4.28}) 
\begin{eqnarray}
&&\dot{v}_{\parallel }^{2}-2\dot{v}_{\parallel }\left\vert \mathbf{\nabla }%
V\right\vert +\mathbf{\dot{v}}_{\perp }^{2}  \notag \\
&=&-\frac{\left( \mathbf{v\nabla }V\right) ^{2}+2\left( \mathbf{v\nabla }%
V\right) \left( \mathbf{v\dot{v}}_{\parallel }\right) +\left( v_{\mathbf{%
\parallel }}\dot{v}_{\parallel }+\mathbf{v}_{\perp }\mathbf{\dot{v}}_{\perp
}\right) ^{2}}{c^{2}-2V-\mathbf{v}^{2}}-\frac{1}{c^{2}}v^{\alpha }v^{\beta
}\partial _{\alpha }\partial _{\beta }V  \label{a4.32}
\end{eqnarray}%
This result distinguishes from the conventional result of the general
relativity, because it depends on the second derivatives $\partial _{\alpha
}\partial _{\beta }V$ of the gravitational potential. Equation (\ref{a4.32})
can be written in the form of quadratic equation with respect to $\dot{v}%
_{\parallel }$ 
\begin{eqnarray}
&&\dot{v}_{\parallel }^{2}\left( 1+\frac{v_{\parallel }^{2}}{c^{2}-2V-%
\mathbf{v}^{2}}\right) -2\dot{v}_{\parallel }\left( \left\vert \mathbf{%
\nabla }V\right\vert -\frac{\left( \mathbf{v\nabla }V\right) v_{\parallel }}{%
c^{2}-2V-\mathbf{v}^{2}}\right) +\left\langle \mathbf{\dot{v}}_{\perp
}^{2}\right\rangle  \notag \\
&=&-\frac{\left( \mathbf{v\nabla }V\right) ^{2}+\left( \mathbf{v}_{\perp }%
\mathbf{\dot{v}}_{\perp }\right) ^{2}}{c^{2}-2V-\mathbf{v}^{2}}-\frac{1}{%
c^{2}}v^{\alpha }v^{\beta }\partial _{\alpha }\partial _{\beta }V
\label{a4.33}
\end{eqnarray}

\section{Fluidity of boundary between the particle \newline
dynamics and space-time geometry}

In the space-time geometry $\mathcal{G}$ the dynamic equations (\ref{b7.2}),
(\ref{b7.3}) are written in the form%
\begin{equation}
\left( \mathbf{P}_{i}^{\left( s\right) }\mathbf{P}_{k}^{\left( s\right) }.%
\mathbf{P}_{i}^{\left( s+1\right) }\mathbf{P}_{k}^{\left( s+1\right)
}\right) =\left\vert \mathbf{P}_{i}^{\left( s\right) }\mathbf{P}_{k}^{\left(
s\right) }\right\vert ^{2},\qquad i,k=0,1,...n  \label{b8.1}
\end{equation}%
\begin{equation}
\left\vert \mathbf{P}_{i}^{\left( s+1\right) }\mathbf{P}_{k}^{\left(
s+1\right) }\right\vert ^{2}=\left\vert \mathbf{P}_{i}^{\left( s\right) }%
\mathbf{P}_{k}^{\left( s\right) }\right\vert ^{2},\qquad i,k=0,1,...n
\label{b8.2}
\end{equation}

The difference dynamic equations (\ref{b8.1}), (\ref{b8.2}) can be written
in the form, which is close to the conventional description in the
Kaluza-Klein space-time \cite{R2008b}. Let $\sigma _{\mathrm{K}_{0}}$ be the
world function in the space-time geometry $\mathcal{G}_{\mathrm{K}_{0}}$.
The geometry $\mathcal{G}_{\mathrm{K}_{0}}$ is the 5D pseudo-Euclidean
geometry of the index 1 with the compactificied coordinate $x^{5}$. In other
words, the space-time geometry $\mathcal{G}_{\mathrm{K}_{0}}$ is the
Kaluza-Klein geometry with vanishing gravitational and electromagnetic
fields. Let us represent the world function $\sigma $ of the space-time
geometry $\mathcal{G}$ in the form 
\begin{equation}
\sigma \left( P,Q\right) =\sigma _{\mathrm{K}_{0}}\left( P,Q\right) +d\left(
P,Q\right)  \label{a5.1}
\end{equation}%
where the function $d$ describes the difference between the true world
function $\sigma $ of the real space-time geometry and the world function $%
\sigma _{\mathrm{K}_{0}}$ of the standard geometry $\mathcal{G}_{\mathrm{K}%
_{0}}$, where the description will be produced. Then one obtains 
\begin{equation}
\left( \mathbf{P}_{0}\mathbf{P}_{1}.\mathbf{Q}_{0}\mathbf{Q}_{1}\right)
=\left( \mathbf{P}_{0}\mathbf{P}_{1}.\mathbf{Q}_{0}\mathbf{Q}_{1}\right) _{%
\mathrm{K}_{0}}+d\left( P_{0},Q_{1}\right) +d\left( P_{1},Q_{0}\right)
-d\left( P_{0},Q_{0}\right) -d\left( P_{1},Q_{1}\right)  \label{a5.2}
\end{equation}%
\begin{equation}
\left\vert \mathbf{P}_{0}\mathbf{P}_{1}\right\vert ^{2}=\left\vert \mathbf{P}%
_{0}\mathbf{P}_{1}\right\vert _{\mathrm{K}_{0}}^{2}+2d\left(
P_{0},P_{1}\right)  \label{a5.3}
\end{equation}%
where index "K$_{0}$" means, that the corresponding quantities are
calculated in the geometry $\mathcal{G}_{\mathrm{K}_{0}}$ by means of the
world function $\sigma _{\mathrm{K}_{0}}$.

By means of (\ref{a5.2}), (\ref{a5.3}) the dynamic equations (\ref{b8.1}), (%
\ref{b8.2}) can be written in the form 
\begin{equation}
\left( \mathbf{P}_{i}^{\left( s\right) }\mathbf{P}_{k}^{\left( s\right) }.%
\mathbf{P}_{i}^{\left( s+1\right) }\mathbf{P}_{k}^{\left( s+1\right)
}\right) _{\mathrm{K}_{0}}-\left\vert \mathbf{P}_{i}^{\left( s\right) }%
\mathbf{P}_{k}^{\left( s\right) }\right\vert _{\mathrm{K}_{0}}^{2}=w\left(
P_{i}^{\left( s\right) },P_{k}^{\left( s\right) },P_{i}^{\left( s+1\right)
},P_{k}^{\left( s+1\right) }\right) ,\qquad i,k=0,1,...n  \label{a5.4}
\end{equation}%
\qquad 
\begin{equation}
\left\vert \mathbf{P}_{i}^{\left( s+1\right) }\mathbf{P}_{k}^{\left(
s+1\right) }\right\vert _{\mathrm{K}_{0}}^{2}-\left\vert \mathbf{P}%
_{i}^{\left( s\right) }\mathbf{P}_{k}^{\left( s\right) }\right\vert _{%
\mathrm{K}_{0}}^{2}=2d\left( P_{i}^{\left( s\right) },P_{k}^{\left( s\right)
}\right) -2d\left( P_{i}^{\left( s+1\right) },P_{k}^{\left( s+1\right)
}\right) ,\qquad i,k=0,1,...n  \label{a5.5}
\end{equation}%
where%
\begin{eqnarray}
w\left( P_{i}^{\left( s\right) },P_{k}^{\left( s\right) },P_{i}^{\left(
s+1\right) },P_{k}^{\left( s+1\right) }\right) &=&2d\left( P_{i}^{\left(
s\right) },P_{k}^{\left( s\right) }\right) -d\left( P_{i}^{\left( s\right)
},P_{k}^{\left( s+1\right) }\right)  \notag \\
&&-d\left( P_{k}^{\left( s\right) },P_{i}^{\left( s+1\right) }\right)
+d\left( P_{i}^{\left( s\right) },P_{i}^{\left( s+1\right) }\right)  \notag
\\
&&+d\left( P_{k}^{\left( s\right) },P_{k}^{\left( s+1\right) }\right)
\label{a5.6}
\end{eqnarray}%
Equations (\ref{a5.4}), (\ref{a5.5}) are dynamic difference equations,
written in the geometry $\mathcal{G}_{\mathrm{K}_{0}}$. Rhs of these
equations can be interpreted as some geometric force fields, generated by
the fact that the space-time geometry $\mathcal{G}$ is described in terms of
some standard geometry $\mathcal{G}_{\mathrm{K}_{0}}$. These force fields
describe deflection of the granular geometry $\mathcal{G}$ from the
Kaluza-Klein geometry $\mathcal{G}_{\mathrm{K}_{0}}$. Such a possibility is
used at the description of the gravitational field, which can be described
as generated by the curvature of the curved space-time, or as a
gravitational field in the space-time geometry of Minkowski. In dynamic
equations (\ref{a5.4}), (\ref{a5.5}) such a possibility is realized for
arbitrary granular space-time geometry $\mathcal{G}$.

Evolution of the leading vector $\mathbf{P}_{0}^{\left( s\right) }\mathbf{P}%
_{1}^{\left( s\right) }$ is of most interest. These equations are obtained
from equations (\ref{a5.4}), (\ref{a5.5}) at $i=0,k=1$. One obtains form
equations (\ref{a5.4}), (\ref{a5.5})%
\begin{equation}
\left\vert \mathbf{P}_{0}^{\left( s+1\right) }\mathbf{P}_{1}^{\left(
s+1\right) }\right\vert _{\mathrm{K}_{0}}^{2}-\left\vert \mathbf{P}%
_{0}^{\left( s\right) }\mathbf{P}_{1}^{\left( s\right) }\right\vert _{%
\mathrm{K}_{0}}^{2}=2d\left( P_{0}^{\left( s\right) },P_{1}^{\left( s\right)
}\right) -2d\left( P_{1}^{\left( s\right) },P_{1}^{\left( s+1\right) }\right)
\label{a5.7}
\end{equation}%
\begin{eqnarray}
&&\left( \mathbf{P}_{0}^{\left( s\right) }\mathbf{P}_{1}^{\left( s\right) }.%
\mathbf{P}_{0}^{\left( s+1\right) }\mathbf{P}_{1}^{\left( s+1\right)
}\right) _{\mathrm{K}_{0}}-\left\vert \mathbf{P}_{0}^{\left( s\right) }%
\mathbf{P}_{1}^{\left( s\right) }\right\vert _{\mathrm{K}_{0}}^{2}  \notag \\
&=&3d\left( P_{0}^{\left( s\right) },P_{1}^{\left( s\right) }\right)
-d\left( P_{0}^{\left( s\right) },P_{1}^{\left( s+1\right) }\right) +d\left(
P_{1}^{\left( s\right) },P_{1}^{\left( s+1\right) }\right)  \label{a5.8}
\end{eqnarray}%
where one uses, that $P_{1}^{\left( s\right) }=P_{0}^{\left( s+1\right) }$.

In the case, when the space-time is uniform, and the function 
\begin{equation}
d\left( P,Q\right) =D\left( \sigma _{\mathrm{K}_{0}}\left( P,Q\right) \right)
\label{a5.9}
\end{equation}%
the equations (\ref{a5.7}), (\ref{a5.8}) take the from%
\begin{equation}
\left\vert \mathbf{P}_{0}^{\left( s+1\right) }\mathbf{P}_{1}^{\left(
s+1\right) }\right\vert _{\mathrm{K}_{0}}^{2}-\left\vert \mathbf{P}%
_{0}^{\left( s\right) }\mathbf{P}_{1}^{\left( s\right) }\right\vert _{%
\mathrm{K}_{0}}^{2}=0  \label{a5.10}
\end{equation}%
\begin{equation}
\left( \mathbf{P}_{0}^{\left( s\right) }\mathbf{P}_{1}^{\left( s\right) }.%
\mathbf{P}_{0}^{\left( s+1\right) }\mathbf{P}_{1}^{\left( s+1\right)
}\right) _{\mathrm{K}_{0}}-\left\vert \mathbf{P}_{0}^{\left( s\right) }%
\mathbf{P}_{1}^{\left( s\right) }\right\vert _{\mathrm{K}_{0}}^{2}=4d\left(
P_{0}^{\left( s\right) },P_{1}^{\left( s\right) }\right) -d\left(
P_{0}^{\left( s\right) },P_{1}^{\left( s+1\right) }\right)  \label{a5.11}
\end{equation}%
In the case, when the leading vector $\mathbf{P}_{0}^{\left( s\right) }%
\mathbf{P}_{1}^{\left( s\right) }$ is timelike, one can introduce the angle $%
\phi _{01}^{\left( s\right) }$ between the vectors $\mathbf{P}_{0}^{\left(
s\right) }\mathbf{P}_{1}^{\left( s\right) }$ and $\mathbf{P}_{0}^{\left(
s+1\right) }\mathbf{P}_{1}^{\left( s+1\right) }$ in the standard geometry $%
\mathcal{G}_{\mathrm{K}_{0}}$. By means of (\ref{a5.10}) it is defined by
the relation 
\begin{equation}
\cosh \phi _{01}^{\left( s\right) }=\frac{\left( \mathbf{P}_{0}^{\left(
s\right) }\mathbf{P}_{1}^{\left( s\right) }.\mathbf{P}_{0}^{\left(
s+1\right) }\mathbf{P}_{1}^{\left( s+1\right) }\right) _{\mathrm{K}_{0}}}{%
\left\vert \mathbf{P}_{0}^{\left( s\right) }\mathbf{P}_{1}^{\left( s\right)
}\right\vert _{\mathrm{K}_{0}}^{2}}  \label{a5.12}
\end{equation}%
Then in the uniform geometry $\mathcal{G}_{\mathrm{K}_{0}}$ the equation (%
\ref{a5.11}) has the form%
\begin{equation}
\sinh \frac{\phi _{01}^{\left( s\right) }}{2}=\frac{\sqrt{4d\left(
P_{0}^{\left( s\right) },P_{1}^{\left( s\right) }\right) -d\left(
P_{0}^{\left( s\right) },P_{1}^{\left( s+1\right) }\right) }}{\sqrt{2}%
\left\vert \mathbf{P}_{0}^{\left( s\right) }\mathbf{P}_{1}^{\left( s\right)
}\right\vert _{\mathrm{K}_{0}}}  \label{a5.14}
\end{equation}%
Thus, relativistic dynamics of particles can be generalized on the case of
the granular space-time geometry.

Application of the world function technique admits one realize the boundary
shift between the particle dynamics and the space-time geometry. A
conceptual development of a theory seems to be more effective in the
monistic conception, having the only basic quantity (world function). The
monistic conception is more sensitive to possible mistakes. This
circumstance admits one to find mistakes and correct them. In the
conception, where there are several basic quantities (concepts), connection
between these quantities may be quite different. This circumstance
embarrasses a choice of correct connection between different basic concepts.
Construction of a conceptual theory, based on physical principles, should be
realized in a form of monistic conception. However, it does not exclude the
fact, that non-monistic theory with a simple space-time geometry may be
simpler in calculations of concrete physical phenomena.

Non-monistic conception is not so sensitive to mistakes in a theory, because
influence of possible mistakes can be compensated in calculations of
concrete physical phenomena by introduction of new hypotheses (and sometimes
by invention of new principles, having a restricted meaning). As far as the
calculations agree with experimental data concerning this physical
phenomenon, the theory is considered as a true theory, confirmed by several
experiments. Such an approach admits one to explain and to calculate new
physical phenomena. However, this approach prevents from construction of
consistent physical theory, which explains all physical phenomena.
Undiscovered mistakes may appear at calculations of other physical phenomena.

Fluidity of boundary between the particle dynamics and the space-time
geometry resolves debate between the adherents of the general relativity and
adherents of the relativistic theory of gravitation.

\section{Pair production}

The pair production effect is considered in the paper on the physics
geometrization, because it was the first appearance of the quantum field
theory (QFT) inconsistency. Impossibility of the conventional QFT to explain
the pair production consistently was the first step, which generated the
idea of the quantum principles revision.

It is considered in the contemporary physics, that the pair production
effect is a special quantum effect, which has no classical analog. It is the
most important evidence in favour of quantum nature of microcosm. However,
it is not so \cite{R2003}. In reality, either classical or quantum mechanism
of pair production is absent in the contemporary quantum field theory, if it
is developed consistently according to quantum principles. Unfortunately,
the quantum field theory was developed inconsistently. Here we describe only
reasons of this inconsistency, referring to original papers for details .

The Klein-Gordon equation (\ref{b3.4}) describes an evolution of a free
quantum object \ (world line). In the absence of electromagnetic field the
equation (\ref{b3.4}) has the form%
\begin{equation}
\hbar ^{2}\partial _{k}\partial ^{k}\psi +m^{2}c^{2}\psi =0  \label{b9.2}
\end{equation}%
Stationary states of this quantum object have the form%
\begin{equation}
\psi =A\exp \left( -ik_{0}t+i\mathbf{kx}\right) ,\qquad k_{0}=\pm \sqrt{%
\mathbf{k}^{2}+\left( \frac{mc}{\hbar }\right) ^{2}}  \label{b9.3}
\end{equation}

If $k_{0}>0$, this quantum object is at the state "particle". If $k_{0}<0$,
this quantum object is at the state "antiparticle". The quantum object is
called "semlon". This term is a reading of the abbreviation "SML", which is
an abbreviation of Russian term "section of world line". Thus, a particle
and an antiparticle are two different states of semlon, but not independent
objects. The semlon has two different states: particle and antiparticle.

Pair production is a generation of a particle and of an antiparticle at some
point of the space-time. The particle and the antiparticle are two different
states of one dynamical system (world line). They cannot be two different
dynamical systems, because two different dynamical objects cannot annihilate
at some space-time point. Effect of pair production or of pair annihilation
appears, when world line changes its direction in the time direction.

Let the world line of a particle be described by the equations 
\begin{equation}
x^{k}=x^{k}\left( \tau \right) ,\qquad k=0,1,2,3  \label{b9.1}
\end{equation}%
where $\tau $ is some parameter (evolution parameter) along the world line.
For the particle $dx^{0}/d\tau >0$. For the antiparticle $dx^{0}/d\tau <0$.
The point, where the derivative $dx^{0}/d\tau $ changes its sign, is a point
of production or a point of annihilation of a pair particle-antiparticle.

The quantity $p_{0}=\hbar k_{0}$ is an eigenvalue of the temporal component $%
\hat{p}_{0}=-i\hbar \partial _{0}$ of the 4-momentum operator 
\begin{equation}
\hat{p}_{k}=-i\hbar \partial _{k}\qquad k=0,1,2,3  \label{b9.4}
\end{equation}%
Particle and antiparticle have different signs of temporal component $%
p_{0}=\hbar k_{0}$ of 4-momentum. The energy $E$ is positive at all semlon
states. 
\begin{equation}
E=\int T_{0}^{0}d\mathbf{x=}\int \left( \hbar ^{2}\left( \partial _{0}\psi
^{\ast }\cdot \partial ^{0}\psi \right) +m^{2}c^{2}\psi ^{\ast }\psi \right)
d\mathbf{x>}0  \label{b9.5}
\end{equation}%
Here $T_{0}^{0}$ is a component of the energy-momentum tensor for the
Klein-Gordon $\mathcal{S}_{\mathrm{KG}}$ dynamic system, having the
Klein-Gordon equation (\ref{b9.2}) as a dynamic equation. Thus, the
evolution operator $\hat{H}=\hat{p}_{0}$ does not coincide in general, with
the energy operator $\hat{E}$, which arises from the expression (\ref{b9.5})
at the second quantization. The same difference takes place at a classical
description of a relativistic particle \cite{R1970}.

In the nonrelativistic approximation, when there is no pair production, the
evolution operator $H$ (Hamiltonian) coincides with the particle energy $E$
(or with $-E$). This coincidence is transmitted to relativistic theory,
where there is a pair production, from nonrelativistic theory, where such a
pair production is absent. The relation 
\begin{equation}
\partial _{0}\psi =\frac{1}{i\hbar }\left[ \psi ,\int T_{0}^{0}d\mathbf{x}%
\right] _{-}  \label{b9.6}
\end{equation}%
where $\left[ ...\right] _{-}$ denotes a commutator, is used in the second
quantized theory for determination of the commutation relations.

These commutation relations lead to consideration of particle and
antiparticle as two different dynamic systems (but not as different states
of the same dynamic system). Formally it means, that the operator $\psi $
contains both creation operators and annihilation operators. It is necessary
to satisfy the relation (\ref{b9.6}). At such a way of the second
quantization the particle and antiparticle are considered as independent
objects. The vacuum state appears to be nonstationary. In general, if the
vacuum state is a state without particles and antiparticles, it has to be
stationary, because in this case the space-time is empty. However, it is
supposed, that the vacuum state contains virtual particles, which may be
converted to real particles and antiparticles, if there is some interaction,
described by nonlinear term, added to the dynamic equation (\ref{b9.2}).

For instance, the pair production appears in the case of nonlinear equation 
\cite{GJ68,GJ70,GJ970,GJ72}%
\begin{equation}
\hbar ^{2}\partial _{k}\partial ^{k}\psi +m^{2}c^{2}\psi =g\psi ^{\ast }\psi
\psi  \label{b9.7}
\end{equation}%
where $g$ is a constant of self-action. Corresponding dynamic equations are
written in form of expansion over the self-action constant $g$. Solving
these equations, one uses a perturbation theory.

There is an alternative representation \cite{R1972} of nonlinear equation (%
\ref{b9.7}), when particle and antiparticle are considered as different
states of a semlon, but not as independent objects. In this case the wave
function $\psi $ contains only annihilation operators, and $\psi ^{\ast }$
contains only creation operators. In this case the energy operator $\hat{E}$
has only nonnegative eigenvalues. The energy $E$ does not coincide with the
Hamiltonian $\hat{H}$, as it takes place in the classical case \cite{R1970}.
The vacuum state is stationary, and there is no necessity for introduction
of virtual particles. The dynamic equations can be written and solved
without expansion over the self-action constant $g$ and without a use of the
perturbation theory. However, in this case there is no pair production.

Absence of the pair production effect means only, that the nonlinear term of
type (\ref{b9.7}) cannot generate the pair production. The pair production
effect is generated by a more complicated interaction, as it follows from (%
\ref{A.1}), (\ref{A.2}), or from \cite{R2003}, where the problem of pair
production is investigated more elaborate. The pair production is connected
with a change of an effective mass $Km$ of a particle, but not with the
virtual particles. It is not clear, how to take into account the change of
effective mass in the framework of quantum theory, although the Klein-Gordon
equation (\ref{b3.4}) takes into account the factor $K$ (\ref{A.2}),
responsible for a change of effective mass.

The problem of pair production is not yet geometrized, although the way of
the pair production geometrization it is rather clear.

\section{Skeleton conception of elementary particles}

After the paper \cite{R91} publication the role of the space-time geometry
increased in the theory of elementary particles, because in fact the quantum
principles were replaced by the multivariant space-time geometry. It became
clear, that constructing a theory of elementary particles, one should use
relativistic concept of the particle state.

In the case, when the particle is not pointlike, its state is described by
its skeleton $\mathcal{P}_{n}=\left\{ P_{0},P_{1},...,P_{n}\right\} $, which
is a set of $(n+1)$ space-time points. These points are connected rigidly.
In the case of a pointlike particle the skeleton consists of two points. The
skeleton $\mathcal{P}_{n}$ is a natural generalization of the skeleton of a
pointlike particle on the case of a composite particle. Motion of any
particle is described by the world chain, consisting of connected skeletons 
\cite{R2008}. ...$\mathcal{P}_{n}^{\left( 0\right) },\mathcal{P}_{n}^{\left(
1\right) },...,\mathcal{P}_{n}^{\left( s\right) }...$%
\begin{equation}
\mathcal{P}_{n}^{\left( s\right) }=\left\{ P_{0}^{\left( s\right)
},P_{1}^{\left( s\right) },..P_{n}^{\left( s\right) }\right\} ,\qquad
s=...0,1,...  \label{b4.8}
\end{equation}%
The adjacent skeletons $\mathcal{P}_{n}^{\left( s\right) },\mathcal{P}%
_{n}^{\left( s+1\right) }$ of the chain are connected by the relations $%
P_{1}^{\left( s\right) }=P_{0}^{\left( s+1\right) }$, $s=...0,1,...$ The
vector $\mathbf{P}_{0}^{\left( s\right) }\mathbf{P}_{1}^{\left( s\right) }=%
\mathbf{P}_{0}^{\left( s\right) }\mathbf{P}_{0}^{\left( s+1\right) }$ is the
leading vector, which determines the world chain direction.

Dynamics of free elementary particle is determined by the relations%
\begin{equation}
\mathcal{P}_{n}^{\left( s\right) }\mathrm{eqv}\mathcal{P}_{n}^{\left(
s+1\right) }:\qquad \mathbf{P}_{i}^{\left( s\right) }\mathbf{P}_{k}^{\left(
s\right) }\mathrm{eqv}\mathbf{P}_{i}^{\left( s+1\right) }\mathbf{P}%
_{k}^{\left( s+1\right) },\qquad i,k=0,1,..n;\qquad s=...0,1,...
\label{a4.8}
\end{equation}%
which describe equivalence of adjacent skeletons. Equivalence of vectors is
defined by the relations (\ref{b7.2a}).

Thus, dynamics of a free elementary particle is described by a system of
algebraic equations (\ref{a4.8}). Specific of dynamics depends on the
elementary particle structure (mutual disposition of points inside the
skeleton) and on the space-time geometry. Lengths $\left\vert \mathbf{P}%
_{i}^{\left( s\right) }\mathbf{P}_{k}^{\left( s\right) }\right\vert $ of
vectors $\mathbf{P}_{i}^{\left( s\right) }\mathbf{P}_{k}^{\left( s\right) }$
are constant along the whole world chain. These $n\left( n+1\right) /2$
quantities may be considered as characteristics of a particle. In the case
of pointlike particle the length $\left\vert \mathbf{P}_{s}\mathbf{P}%
_{s+1}\right\vert $ of the link $\mathbf{P}_{s}\mathbf{P}_{s+1}$ is the
geometrical mass of the particle. In the case of a more complicated
skeletons the meaning of parameters $\left\vert \mathbf{P}_{i}^{\left(
s\right) }\mathbf{P}_{k}^{\left( s\right) }\right\vert $ should be
investigated.

The system of dynamic equations (\ref{a4.8}) consists of $n\left( n+1\right) 
$ algebraic equations for $nD$ dynamic variables, where $D$ is the dimension
of the space-time (the number of coordinates, which are necessary for
labeling of all points of the space-time). If $n\leq D$, the number of
dynamic variables is less, than the number of dynamic equations. In this
case we have a discrimination mechanism, which forbids some skeletons. This
mechanism admits one to explain discrete parameters of elementary particles.
If $n>D+1$, the number of dynamic equations is more than the number of
dynamic variables. In this case there may exist many solutions, and the
particle motion becomes multivariant (stochastic). Both cases may take place
in the theory of elementary particles.

Dynamic equations (\ref{a4.8}) are written in the coordinateless form, and
this fact is a worth of the dynamic equations (\ref{a4.8}), as far as it
saves from a necessity to consider the coordinate transformations. Dynamic
equations (\ref{a4.8}) are algebraic equations (not differential), and this
fact is also a worth of the theory, because the algebraic equations are
insensitive to a possible discreteness of the space-time geometry.

The first (nontrivial) attempt of a use of the relativistic concept of the
particle state was made. One considered the structure of the Dirac particle
(fermion) \cite{R2008a}. It was a conceptual step, because a possibility of
spacelike world chains was considered. Spacelike world lines of real
particles are absent in the conventional conception of elementary particle
(they are possible for virtual particles, but it is a special problem). In
the skeleton conception of elementary particles such a restriction is absent.

The Dirac particle is a dynamic system $\mathcal{S}_{\mathrm{D}}$, whose
dynamic equation is the Dirac equation 
\begin{equation}
i\gamma ^{k}\partial _{k}\psi +mc\psi =0  \label{b10.1}
\end{equation}%
It appeared that the skeleton of the Dirac particle consists of $n$ points ($%
n\geq 3$). Its world chain is a spacelike helix with a timelike axis.

In our calculations we used the mathematical technique \cite{S930,S51},
where $\gamma $-matrices are represented as hypercomplex numbers. Using
designations 
\begin{equation}
\gamma _{5}=\gamma ^{0123}\equiv \gamma ^{0}\gamma ^{1}\gamma ^{2}\gamma
^{3},  \label{f1.9}
\end{equation}%
\begin{equation}
\mathbf{\sigma }=\{\sigma _{1},\sigma _{2},\sigma _{3},\}=\{-i\gamma
^{2}\gamma ^{3},-i\gamma ^{3}\gamma ^{1},-i\gamma ^{1}\gamma ^{2}\}
\label{f1.10}
\end{equation}%
we make the change of variables 
\begin{equation}
\psi =Ae^{i\varphi +{\frac{1}{2}}\gamma _{5}\kappa }\exp \left( -\frac{i}{2}%
\gamma _{5}\mathbf{\sigma \eta }\right) \exp \left( {\frac{i\pi }{2}}\mathbf{%
\sigma n}\right) \Pi   \label{f1.11}
\end{equation}%
\begin{equation}
\psi ^{\ast }=A\Pi \exp \left( -{\frac{i\pi }{2}}\mathbf{\sigma n}\right)
\exp \left( -\frac{i}{2}\gamma _{5}\mathbf{\sigma \eta }\right) e^{-i\varphi
-{\frac{1}{2}}\gamma _{5}\kappa }  \label{f1.12}
\end{equation}%
where (*) means the Hermitian conjugation, and 
\begin{equation}
\Pi ={\frac{1}{4}}(1+\gamma ^{0})(1+\mathbf{z\sigma }),\qquad \mathbf{z}%
=\{z^{\alpha }\}=\text{const},\qquad \alpha =1,2,3;\qquad \mathbf{z}^{2}=1
\label{f1.13}
\end{equation}%
is a zero divisor. The quantities $A$, $\kappa $, $\varphi $, $\mathbf{\eta }%
=\{\eta ^{\alpha }\}$, $\mathbf{n}=\{n^{\alpha }\}$, $\alpha =1,2,3,\;$ $%
\mathbf{n}^{2}=1$ are eight real parameters, determining the wave function $%
\psi .$ These parameters may be considered as new dependent variables,
describing the state of dynamic system $\mathcal{S}_{\mathrm{D}}$. The
quantity $\varphi $ is a scalar, and $\kappa $ is a pseudoscalar. Six
remaining variables $A,$ $\mathbf{\eta }=\{\eta ^{\alpha }\}$, $\mathbf{n}%
=\{n^{\alpha }\}$, $\alpha =1,2,3,\;$ $\mathbf{n}^{2}=1$ can be expressed
through the flux 4-vector 
\begin{equation}
j^{l}=\bar{\psi}\gamma ^{l}\psi ,\qquad l=0,1,2,3  \label{f1.13b}
\end{equation}%
and spin 4-pseudovector 
\begin{equation}
S^{l}=i\bar{\psi}\gamma _{5}\gamma ^{l}\psi ,\qquad l=0,1,2,3  \label{f1.13a}
\end{equation}%
Because of two identities 
\begin{equation}
S^{l}S_{l}\equiv -j^{l}j_{l},\qquad j^{l}S_{l}\equiv 0.  \label{f1.14}
\end{equation}%
there are only six independent components among eight components of
quantities $j^{l}$, and $S^{l}$. Now we can write the action for the dynamic
equation(\ref{b10.1}) in the hydrodynamical form 
\begin{equation}
\mathcal{S}_{\mathrm{D}}:\qquad \mathcal{A}_{D}[j,\varphi ,\kappa ,\mathbf{%
\xi }]=\int \mathcal{L}d^{4}x,\qquad \mathcal{L}=\mathcal{L}_{\mathrm{cl}}+%
\mathcal{L}_{\mathrm{q1}}+\mathcal{L}_{\mathrm{q2}}  \label{c4.15}
\end{equation}%
\begin{equation}
\mathcal{L}_{\mathrm{cl}}=-m\rho -\hbar j^{i}\partial _{i}\varphi -\frac{%
\hbar j^{l}}{2\left( 1+\mathbf{\xi z}\right) }\varepsilon _{\alpha \beta
\gamma }\xi ^{\alpha }\partial _{l}\xi ^{\beta }z^{\gamma },\qquad \rho
\equiv \sqrt{j^{l}j_{l}}  \label{c4.16}
\end{equation}%
\begin{equation}
\mathcal{L}_{\mathrm{q1}}=2m\rho \sin ^{2}(\frac{\kappa }{2})-{\frac{\hbar }{%
2}}S^{l}\partial _{l}\kappa ,  \label{c4.17}
\end{equation}%
\begin{equation}
\mathcal{L}_{\mathrm{q2}}=\frac{\hbar (\rho +j_{0})}{2}\varepsilon _{\alpha
\beta \gamma }\partial ^{\alpha }\frac{j^{\beta }}{(j^{0}+\rho )}\xi
^{\gamma }-\frac{\hbar }{2(\rho +j_{0})}\varepsilon _{\alpha \beta \gamma
}\left( \partial ^{0}j^{\beta }\right) j^{\alpha }\xi ^{\gamma }
\label{c4.18}
\end{equation}%
Lagrangian is a function of 4-vector $j^{l}$, scalar $\varphi $,
pseudoscalar $\kappa $, and unit 3-pseudovector $\mathbf{\xi }$, which is
connected with the spin 4-pseudovector $S^{l}$ by means of the relations 
\begin{equation}
\xi ^{\alpha }=\rho ^{-1}\left[ S^{\alpha }-\frac{j^{\alpha }S^{0}}{%
(j^{0}+\rho )}\right] ,\qquad \alpha =1,2,3;\qquad \rho \equiv \sqrt{%
j^{l}j_{l}}  \label{f1.15}
\end{equation}%
\begin{equation}
S^{0}=\mathbf{j\xi },\qquad S^{\alpha }=\rho \xi ^{\alpha }+\frac{(\mathbf{%
j\xi })j^{\alpha }}{\rho +j^{0}},\qquad \alpha =1,2,3  \label{f1.16}
\end{equation}%
\begin{equation*}
\end{equation*}%
Let us produce dynamical disquantization \cite{R2001,R2005} of the action (%
\ref{c4.15})--(\ref{c4.18}), making the change 
\begin{equation}
\partial _{k}\rightarrow \frac{j_{k}j^{s}}{j^{l}j_{l}}\partial _{s}
\label{f1.17}
\end{equation}%
The action (\ref{c4.15})--(\ref{c4.18}) takes the form 
\begin{eqnarray}
\mathcal{A}_{\mathrm{Dqu}}[j,\varphi ,\kappa ,\mathbf{\xi }] &=&\int \left\{
-m\rho \cos \kappa -\hbar j^{i}\left( \partial _{i}\varphi +\frac{%
\varepsilon _{\alpha \beta \gamma }\xi ^{\alpha }\partial _{i}\xi ^{\beta
}z^{\gamma }}{2\left( 1+\mathbf{\xi z}\right) }\right) \right.   \notag \\
&&+\left. \frac{\hbar j^{k}}{2(\rho +j_{0})\rho }\varepsilon _{\alpha \beta
\gamma }\left( \partial _{k}j^{\beta }\right) j^{\alpha }\xi ^{\gamma
}\right\} d^{4}x  \label{f1.18}
\end{eqnarray}%
Note that the second term $-\frac{\hbar }{2}S^{l}\partial _{l}\kappa $ in
the relation (\ref{c4.17}) is neglected, because 4-pseudovector $S^{k}$ is
orthogonal to 4-vector $j^{k}$, and the derivative $S^{l}\partial
_{||l}\kappa =S^{l}\rho ^{-2}j_{l}j^{k}\partial _{k}\kappa $ vanishes.

Although the action (\ref{f1.18}) contains a non-classical variable $\kappa $%
, in fact this variable is a constant. Indeed, a variation with respect to $%
\kappa $ leads to the dynamic equation 
\begin{equation}
\frac{\delta \mathcal{A}_{Dqu}}{\delta \kappa }=m\rho \sin \kappa =0
\label{f1.19}
\end{equation}%
which has solutions 
\begin{equation}
\kappa =n\pi ,\qquad n=\func{integer}  \label{f1.20}
\end{equation}%
Thus, the effective mass $m_{\mathrm{eff}}=m\cos \kappa $ has two values 
\begin{equation}
m_{\mathrm{eff}}=m\cos \kappa =\kappa _{0}m=\pm m  \label{f1.21}
\end{equation}%
where $\kappa _{0}$ is a dichotomic quantity $\kappa _{0}=\pm 1$ introduced
instead of cos $\kappa $. The quantity $\kappa _{0}$ is a parameter of the
dynamic system $\mathcal{S}_{\mathrm{Dqu}}$. It is not to be varying. The
action (\ref{f1.18}), turns into the action 
\begin{eqnarray}
\mathcal{A}_{\mathrm{Dqu}}[j,\varphi ,\mathbf{\xi }] &=&\int \left\{ -\kappa
_{0}m\rho -\hbar j^{i}\left( \partial _{i}\varphi +\frac{\varepsilon
_{\alpha \beta \gamma }\xi ^{\alpha }\partial _{i}\xi ^{\beta }z^{\gamma }}{%
2\left( 1+\mathbf{\xi z}\right) }\right) \right.   \notag \\
&&+\left. \frac{\hbar j^{k}}{2(\rho +j_{0})\rho }\varepsilon _{\alpha \beta
\gamma }\left( \partial _{k}j^{\beta }\right) j^{\alpha }\xi ^{\gamma
}\right\} d^{4}x  \label{f1.22}
\end{eqnarray}

Let us introduce Lagrangian coordinates $\tau =\{\tau _{0},\mathbf{\tau }%
\}=\{\tau _{i}\left( x\right) \}$, $i=0,1,2,3$ as functions of coordinates $x
$ in such a way that only coordinate $\tau _{0}$ changes along the direction 
$j^{l}$. The action (\ref{f1.22}) is transformed to the form 
\begin{equation}
\mathcal{A}_{\mathrm{Dqu}}[x,\mathbf{\xi }]=\int \mathcal{A}_{\mathrm{Dcl}%
}[x,\mathbf{\xi }]d\mathbf{\tau ,\qquad d\tau }=d\tau _{1}d\tau _{2}d\tau
_{3}  \label{f1.23}
\end{equation}%
where 
\begin{equation}
\mathcal{S}_{\mathrm{Dcl}}:\qquad \mathcal{A}_{\mathrm{Dcl}}[x,\mathbf{\xi }%
]=\int \left\{ -\kappa _{0}m\sqrt{\dot{x}^{i}\dot{x}_{i}}+\hbar {\frac{(\dot{%
\mathbf{\xi }}\times \mathbf{\xi })\mathbf{z}}{2(1+\mathbf{\xi z})}}+\hbar 
\frac{(\dot{\mathbf{x}}\times \ddot{\mathbf{x}})\mathbf{\xi }}{2\sqrt{\dot{x}%
^{s}\dot{x}_{s}}(\sqrt{\dot{x}^{s}\dot{x}_{s}}+\dot{x}^{0})}\right\} d\tau
_{0}  \label{f1.24}
\end{equation}%
After dynamic disquantization the Dirac particle is a statistical ensemble
of dynamic systems $\mathcal{S}_{\mathrm{Dcl}}$, as it follows from (\ref%
{f1.23}) and (\ref{f1.24}). Any dynamic system $\mathcal{S}_{\mathrm{Dcl}}$
has $10$ degrees of freedom. 6 degrees of freedom describe a progressive
motion of a particle and 4 degrees of freedom describe the rotational motion
of the particle. It is a classical model of the Dirac particle, which
contains the quantum constant. The quantum constant appears in classical
dynamic equations, because these equations are to contain magnetic moment.
But the magnetic moment, (classical quantity!) depends on the quantum
constant.

The variables $\mathbf{\xi }$ describe rotation, which is a classical analog
of so-called "zitterbewegung". The Dirac particle is not a pointlike
particle \cite{R2004}. Description of internal degrees of freedom in terms
of $\mathbf{\xi }$ appears to be nonrelativistic \cite{R2003a,R2004b},
although the translational degrees of freedom $x$ are described
relativistically.

One succeeds to describe the classical model $\mathcal{S}_{\mathrm{Dcl}}$ of
Dirac particle in the framework of the skeleton conception of the elementary
particles. The discrete space-time geometry (\ref{b1.3}) is replaced by
half-discrete space-time geometry, described by the world function $\sigma _{%
\mathrm{d}}$%
\begin{equation}
\sigma _{\mathrm{d}}=\sigma _{\mathrm{M}}+\lambda _{0}^{2}\left\{ 
\begin{array}{lll}
\mathrm{sgn}\left( \sigma _{\mathrm{M}}\right) & \text{if} & \left\vert
\sigma _{\mathrm{M}}\right\vert >\sigma _{0}>0 \\ 
f\left( \sigma _{\mathrm{M}}\right) & \text{if} & \left\vert \text{\ }\sigma
_{\mathrm{M}}\right\vert <\sigma _{0}%
\end{array}%
\right. \qquad \lambda _{0}^{2}=\frac{\hbar }{2bc}  \label{a4.14}
\end{equation}%
where $\sigma _{\mathrm{M}}$ is the world function of the space-time of
Minkowski, $b$ is some universal constant and $\sigma _{0}$ is some
constant. The function $f$ is a monotone nondecreasing function, having
properties $f\left( -\sigma _{0}\right) =-1$, $f\left( \sigma _{0}\right) =1$%
.

The space-time geometry, described by the world function (\ref{a4.14}) is
uniform and isotropic. The part of the world function corresponding to $%
\left\vert \sigma _{\mathrm{M}}\right\vert >\sigma _{0}$ is responsible for
quantum effects of pointlike particle (Schr\"{o}dinger equation \cite{R91}).
The part of the world function (\ref{a4.14}), corresponding to $\left\vert
\sigma _{\mathrm{M}}\right\vert <\sigma _{0}$ is responsible for the
structure of a particle with the skeleton consisting of more, than two
points. If $\left\vert f\left( \sigma _{\mathrm{M}}\right) \right\vert
<\left\vert \sigma _{\mathrm{M}}/\sigma _{0}\right\vert $, the spacelike
world chain may have a shape of a helix with a timelike axis.

The case, when 
\begin{equation}
f\left( \sigma _{\mathrm{M}}\right) =\left( \frac{\sigma _{\mathrm{M}}}{%
\sigma _{0}}\right) ^{3}  \label{a2.5}
\end{equation}%
has been investigated \cite{R2008a}. Such a choice of the world function
does not pretend to description of the real space-time. It is only some
model, which correctly describes quantum effects connected with pointlike
particles and tries to investigate, whether spacelike world chain may have a
shape of a helix with a timelike axis. According to semiclassical
approximation of the Dirac equation \cite{R1995,R2004,R2004b} the world line
of a \textit{free classical} Dirac particle has the shape of a helix. Such a
shape of the world line explains existence of a spin. It was interesting,
whether the spin of the Dirac particle can be obtained in the skeleton
conception of elementary particles.

Consideration in \cite{R2008a} confirmed the supposition on the helix world
chain of the Dirac particle (fermion). The skeleton of a fermion is to
contain more, than two points. Besides, some restrictions on disposition of
the skeleton points were obtained. It means that in the skeleton conception
there is a discrimination mechanism responsible for discrete values of
parameters of the elementary particles. Such a discrimination mechanism is
absent in the conventional approach, based on a use of quantum principles.
The obtained results are preliminary, because the simple restriction (\ref%
{a2.5}) on the world function has been used. Nevertheless these results
show, that the skeleton conception admits one to investigate the structure
of elementary particles. 

The conventional approach, based on quantum principles, admits one only to
ascribe to elementary particles such phenomenological properties as mass,
spin, color, flavour and other, without explanation how these properties
relate to the elementary particle structure. Conventional approach admits
one only to classify elementary particles by their phenomenological
properties and to predict reaction between the elementary particles on the
basis of this classification.

Such a situation reminds situation with investigation of chemical elements.
Periodic system of chemical elements is a phenomenological construction. It
is an attribute of chemistry. Arrangement of atoms of chemical elements is
investigated by physics (quantum mechanics). The periodic system of chemical
elements had been discovered earlier, than one began to investigate atomic
structures. However, the periodic system did not help us to create quantum
mechanics and to investigate the atomic structure. The periodic system and
the quantum mechanics are attributes of different sciences. In the same
track the skeleton conception of elementary particles and the conventional
phenomenological approach are essentially attributes of different sciences,
investigating different sides of the elementary particles.

\section{Conclusions}

Thus, in the twentieth century a transition from the nonrelativistic physics
to the relativistic one has been produced only in dynamic equations, but not
in the concept of the particle state. The particle state as a point of the
phase space is inadequate in application to indeterministic particles. In
the nonrelativistic physics the particle state is described as a point in
the phase space. Existence of primordially indeterministic particles in the
microcosm does not admit a use of phase space, because the limit (\ref{b1.7}%
), determining the particle momentum, does not exist for indeterministic
particles. One is forced to describe the particle state without limits of
the type (\ref{b1.7}).

The relativistic concept of the particle state is realized by means of the
particle skeleton. The skeleton consists of several space-time points. Such
a concept of the particle state can be applied both for deterministic and
indeterministic particles. The number of the skeleton points depends on the
structure of the elementary particle. It is important, that the skeleton
describes all characteristics of the particle, including its mass, charge,
momentum and other characteristics, if they take place, (spin, flower, etc.
). As a result one obtains a monistic conception, where all fundamental
physical phenomena (including electromagnetic and gravitational
interactions) are described in terms of points of the event space and of
world functions between them.

Dynamic equations are algebraic equations, formulated in the coordinateless
form. These equations are simpler and more universal, than equations, used
in conventional theory of elementary particles.

The obtained skeleton conception is not yet a theory of elementary
particles. It is only a conception, which deals with physical and
geometrical principles. It is supposed that the skeleton conception can be
applied for any space-time geometry and for any skeletons, which are
compatible with this space-time geometry. In reality, there is a real
space-time geometry, and there are only those skeletons which are admitted
by this space-time geometry. The skeleton conception turns to a theory of
elementary particles, only when this real space-time geometry will be
determined. This real space-time geometry and skeletons, which are
compatible with this geometry are to agree with experimental data.

The conception (physical principles) $\mathcal{C}_{\mathrm{con}}$ of the
conventional theory of elementary particles is inconsistent, because it uses
nonrelativistic concept of the particle state, which cannot be used at
description of indeterministic particles. Any inconsistent theory has a very
useful property. \textit{Such a theory admits one to obtain any desirable
statement. One needs only to invent a proper hypothesis}. A consistent
theory admits one to obtain only those statements, which follow from basic
statements of the theory, even if anybody wants to obtain another
statements. The consistent theory admits one to introduce only those
additional hypotheses, which are compatible with the theory.

Experimenters investigating elementary particle need  some concepts for
description of their experiments. They cannot describe their experiments
without a use of some concepts. The experimenters take these concepts from
their experience and from existing theories. Unfortunately, these concepts
are adequate not always. System of these concepts is phenomenological. It
useful for description of experiments. However, it is not always adequate
for description of the elementary particles nature. One can see this in the
example of atomic structure investigation. Chemists, who investigated
experimentally properties of chemical elements, knew nothing on the atoms
arrangement. The atom structure cannot be described in those
phenomenological concepts, which were used by chemists.

The presented skeleton conception is only a conception (but not a theory) of
elementary particles. It cannot be tested experimentally. One needs to
determine a real space-time geometry and to investigate possible skeletons
of elementary particles. Then the skeleton conception turns to a theory of
elementary particles, and it can be tested experimentally\label{0en}

\end{document}